\documentclass[prd,nofootinbib,preprintnumbers]{revtex4-2}
\usepackage{epsfig,float}
\usepackage{amsfonts}
\usepackage{amsmath}
\usepackage{longtable}
\usepackage[colorlinks=true,linkcolor=red,citecolor=blue]{hyperref}
\usepackage{wrapfig}
\usepackage{enumitem}
\usepackage{nicefrac}
\usepackage[parfill]{parskip}
\usepackage{orcidlink}
\newcommand{\im}{{\rm Im}}

\newcommand{\be}{\begin{eqnarray}}
\newcommand{\ee}{\end{eqnarray}}
\newcommand{\ba}{\begin{array}}
\newcommand{\ea}{\end{array}}

\newcommand{\bea}{\begin{eqnarray}}
\newcommand{\eea
}{\end{eqnarray}}
\newcommand{\bi}{\begin{itemize}}
\newcommand{\ei}{\end{itemize}}
\newcommand{\nn}{{\nonumber}}

\restylefloat{figure}
\restylefloat{table}

\begin{document}

\title{Systematic description of hadron’s response to non-local QCD probes: Froissart-Gribov projections in analysis of deeply virtual Compton scattering}

\author{Kirill~M.~Semenov-Tian-Shansky$^{1,2,3}$\orcidlink{0000-0001-8159-0900},
Paweł Sznajder$^{4}$\orcidlink{0000-0002-2684-803X}}
\affiliation{
$^1$ Department of Physics, Kyungpook National University, Daegu 41566, Korea \\
$^2$ National Research Centre Kurchatov Institute: Petersburg Nuclear Physics Institute, 188300 Gatchina, Russia \\
$^3$ Higher School of Economics,
National Research University, 194100 St. Petersburg, Russia \\
$^4$ National Centre for Nuclear Research, NCBJ, 02-093 Warsaw, Poland
}

\begin{abstract}
We revisit the application of the Froissart-Gribov (FG) projections in the analysis of amplitudes for Deeply Virtual Compton Scattering (DVCS), providing essential information on generalized parton distributions (GPDs). The pivotal role of these projections in a systematic description of a hadron's response to the string-like QCD probes characterised by different values of angular momentum $J$ is emphasized. For the first time, we establish a relationship between the FG projections and GPDs for
spin-\nicefrac{1}{2} targets, and we investigate these quantities in various GPD models. Finally, we provide the first numerical estimates for the FG projections based on the DVCS amplitudes directly extracted from experimental data.
We argue  the method of the FG projections deserves a broad application in the DVCS phenomenology.
\end{abstract}

\maketitle
\thispagestyle{empty}
\renewcommand{\thesection}{\arabic{section}}
\renewcommand{\thesubsection}{\arabic{subsection}}

\section{Introduction}

Generalized parton distributions (GPDs)
\cite{Mueller:1998fv,Radyushkin:1997ki,Ji:1996nm,Ji:1998pc}
are recognized as a convenient tool to address
the dynamics of hadron constituents and to provide a QCD-based description of internal structure of
hadrons (see Refs.~\cite{Goeke:2001tz,Diehl:2003ny,Belitsky:2005qn,Boffi:2007yc}
for a review).
According to the QCD collinear factorization theorem
\cite{Ji:1996nm},
GPDs encode the response of a hadronic target to an excitation induced by the well-defined
QCD string quark and gluon operators on the light-cone $z^2=0$:
\begin{eqnarray}
\label{AtoB}
&&\langle N'| \bar \psi (0) [0; \,z] \psi (z) |N\rangle \,, \nonumber \\
&&\langle N'| G_{\alpha \beta}^a(0)\ [0,z]^{ab}\ G_{\mu \nu}^{b}(z) |N\rangle\,,
\label{QCD_String_Operators}
\end{eqnarray}
where $[0;\,z]$ ($ [0,z]^{ab}$)  stand for the Wilson lines in the fundamental (adjoint) representations ensuring the gauge invariance of the quark (gluon) operators.

The key advantage for investigating hadronic structures in hard exclusive reactions admitting a description within the GPD framework
is brought by non-local nature of the QCD string operators (\ref{QCD_String_Operators}). This allows to enlarge the inventory of  ``elementary'' probes available for
resolving the hadron constituents. Namely, the truly elementary probes are only limited to spin $J=1$
photons ($\gamma$) and  heavy vector bosons ($W^\pm, \, Z^0$),
while the GPD framework allows for the emulation of the $J=2$ graviton probe. This unique feature gives access to hadronic gravitational form factors (GFFs), that are defined from  hadronic matrix elements of QCD energy-momentum tensor (EMT).
Study of hadronic GFFs has provided new tools to address the spin contents of hadrons
relying on Ji's sum rule
\cite{Ji:1996ek}
and to investigate the mechanical properties of hadronic medium encoded in the $D$-term
form factor
\cite{Polyakov:1998ze, Polyakov:2018zvc}.
First experimental extractions of the pressure distribution in a proton
presented in \cite{Burkert:2018bqq,Kumericki:2019ddg,Dutrieux:2021nlz,Burkert:2021ith} demonstrated the
robustness of the approach
and initiated proposals for dedicated studies with perspective
hadronic experimental facilities, such as the upgraded JLab@20~GeV \cite{Accardi:2023chb}, J-PARC \cite{Aoki:2021cqa}; and the
EIC \cite{AbdulKhalek:2021gbh,Burkert:2022hjz,Abir:2023fpo}, see {\it e.g.} Ref.~\cite{Diehl:2023nmm} for an overview.

Hard exclusive reactions admitting a description in terms of GPDs, such as the Deeply Virtual Compton Scattering (DVCS) and Deeply Virtual Meson Production (DVMP), only provide information on convolutions of GPDs with hard scattering kernels,  giving rise to the broadly debated deconvolution problem, see discussion {\it e.g.} in Refs. \cite{Kumericki:2008di,Bertone:2021yyz}. In case of the DVCS, the maximum information on GPDs one may expect to extract from experiments performed for a fixed photon virtuality $Q^2$
turns to be limited to GPDs on so-called cross-over trajectory $x=\xi$ and the $D$-term form factor, the subtraction constant appearing in a once-subtracted fixed-$t$ dispersion relation for the Compton amplitude. Additional information can be obtained accounting for the QCD evolution effects provided a sufficiently
large lever arm in $Q^2$ \cite{Freund:1999xf,Moffat:2023svr,Dutrieux:2021nlz}. Direct access to GPDs beyond the $x=\xi$ trajectory can be provided by Double Deeply Virtual Compton Scattering (DDVCS) \cite{Belitsky:2002tf, Guidal:2002kt, Deja:2023ahc}, however this process turns to be highly challenging to measure.

The cross channel SO$(3)$ partial waves (PWs) labeled by the cross channel angular momentum $J$ were introduced
in
Ref.~\cite{Kumericki:2007sa}
in the framework of the
conformal PW expansion of GPDs
as a convenient and physically motivated basis for expanding the conformal moments of GPDs.
The equivalence between the double (conformal and cross channel SO$(3)$) PW expansion and the the dual parametrization of GPDs \cite{Polyakov:2002wz}
was finally elaborated in Ref.~\cite{Muller:2014wxa}.
The Froissart-Gribov (FG) projection  \cite{Gribov:1961ex,Froissart:1961ux} was first addressed in the
context of the DVCS in Ref.~\cite{Kumericki:2008di}. That study recognized the FG projections of the Compton Form Factors (CFFs) as a source of strong constraints for GPDs through the GPD sum rules.
In particular, the FG projections of CFFs make a direct contact with the extension of the Regge phenomenology to the off-shell kinematics, specific for DVCS, through the analysis of the $t$-dependent singularities of PWs in the complex-$J$ plane.

\begin{figure}[H]
\label{Fig_J}
\begin{center}
\includegraphics[width=0.40\textwidth]{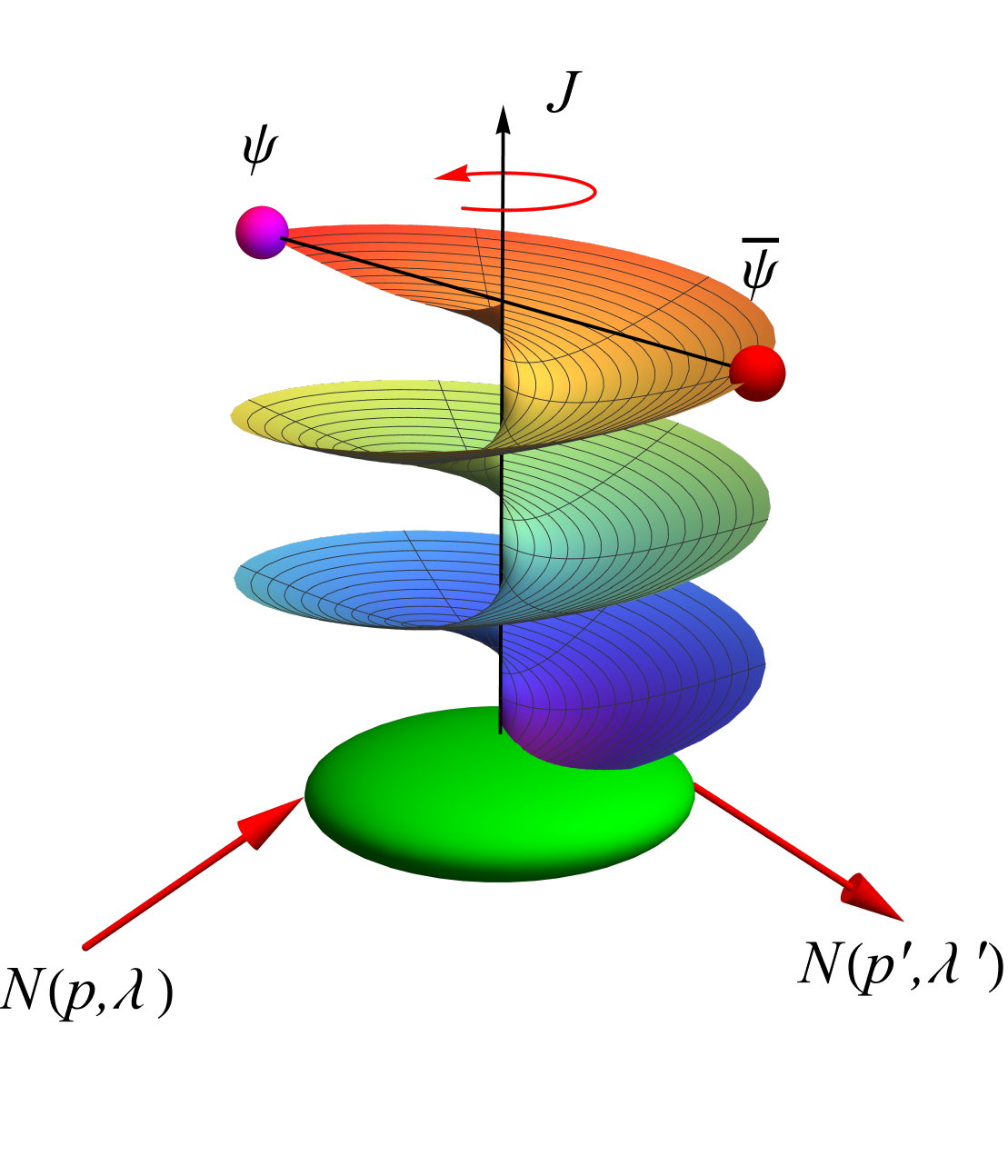}
\end{center}
\caption{A non-local QCD quark probe provides a tower of local operators of spin-$J$ with invariant momentum transfer $\Delta^2=(p'-p)^2$ exciting the target nucleon.}
\end{figure}

The same issue was also considered from a slightly different
perspective within the Abel transform tomography analysis
\cite{Polyakov:2007rw,Polyakov:2007rv}
of the DVCS in the framework of the dual parametrization of GPDs.
The Mellin moments of the so-called ``GPD quintessence function'' $N$
recovered from the absorptive part of the CFF  by the inverse Abel transformation were found to quantify the target hadron’s response on the cross channel excitation with a fixed angular momentum $J$.
This provides an appealing possibility to decompose the string-like QCD probes
(\ref{QCD_String_Operators})
into a tower of cross channel excitations of certain $J$, and may open
new possibilities for studies of hadronic structure focusing on the hadron
target response on a probe with a specific cross channel angular momentum $J$, see Fig.~\ref{Fig_J}.
By extending the formalism to the case of non-diagonal DVCS and DVMP
reactions (see {\it e.g.} \cite{Semenov-Tian-Shansky:2023bsy,Kroll:2022roq}) one may develop new tools for hadron spectroscopy
studying resonance production by means of a probe with a selected value of $J$.

In this paper we revisit the application of the FG projections in the context of the DVCS.
The paper is organized as follows. We start with specifying our conventions and notations in Sect.~\ref{Sec_Conventions}.
By following the exposition of Ref.~\cite{Muller:2014wxa}, in Sect.~\ref{Sec_Spinless_case} we review the derivation of the FG projection of the CFF for the case of spinless target hadron. In addition, we inspect the connection between the FG projections and the Abel tomography method in the dual parametrization framework. We establish a set of constructive sum rules that connect the FG projections of the CFF
to coefficients at powers of $\xi$ of the Mellin moments of GPDs. The latter can be computed within the
phenomenological GPD models, and can be studied with the methods of lattice QCD through the correspondence with the
form factors occurring in the decomposition of hadronic matrix elements of local quark
twist-$2$ operators.
We also present a discussion of
mixing of the cross channel SO$(3)$ PWs due to non-zero target mass corrections.
In Sect.~\ref{Sec_Nucleon_target} we extend the FG projection framework for
the case of spin-$\nicefrac{1}{2}$ target. While the case of electric combination of nucleon CFFs,
${\cal H}^{(E)} \equiv {\cal H}+ \frac{t}{4m^2} {\cal E}$,
is fully analogous to the spinless target case,
we work out the FG projection for the magnetic combination of
CFFs ${\cal H}^{(M)} \equiv {\cal H}+   {\cal E}$.
We present its interpretation from the perspective of the Abel transform tomography
framework within the dual parametrization approach and work out the sum rules for the nucleon
CFF FG projections.
In Sect.~\ref{Sec_Experiment_and_Models} we address the phenomenological application of the
FG projections of the CFFs.
We compute several first FG projections of both electric and magnetic combinations
${\cal H}^{(E,M)}$
of CFFs using the
Goloskokov-Kroll (GK)
\cite{Goloskokov:2005sd,Goloskokov:2008ib},
Kumeri\v{c}ki-M\"{u}ller (KM) \cite{Kumericki:2015lhb} and
Mezrag-Moutarde-Sabati\'e (MMS) \cite{Mezrag:2013mya}
phenomenological GPD models presently employed
for the analysis of the DVCS data (for the GK see Ref.~\cite{Kroll:2012sm}).
The FG projections turn to be rather discriminating between the
GPD models; and we expect the method to see a broad application in the DVCS phenomenology.
We also compare the model results for the FG projections with those obtained from the
global model-independent extraction of the DVCS CFFs of
Ref.~\cite{Moutarde:2019tqa}.
At the moment the overall incertitude of the method  remains too large to clearly
distinguish between the phenomenological models. However, for the FG projections of the better known
electric combination of CFFs, the magnitude  of an error shows a tendency to reduce in a
certain range of $t$. Therefore,
with the new precise experimental data on DVCS one may expect the method will ultimately  gain a discriminative power. Finally, Sect.~\ref{Sec_Conclusions} presents our Conclusions and possible Outlook.

\section{Conventions and notations}
\label{Sec_Conventions}

Our system of conventions mainly follows Ref.~\cite{Diehl:2003ny}.
GPDs are functions of three variables: $x=k^{+}/P^{+}$ describing the average longitudinal momentum of the active parton, the skewness variable $\xi=(p^{+}-p'^{+})/(p^{+}+p'^{+})=-\Delta^{+}/(2P^{+})$ characterising the longitudinal momentum transfer, and the invariant momentum transfer variable $t=\Delta^2$. Here, $P \equiv \frac{p'+p}{2}$ is the average momentum; and $\Delta \equiv p'-p$ is the momentum transfer between the final and initial state hadrons, while $k-\Delta/2$ and $k+\Delta/2$ are the four-momenta of the emitted and reabsorbed partons, respectively. We employ the following light-cone vectors $n_{\pm}=(1,0,0,\pm 1)/\scriptstyle\sqrt{2}$, such that for a given four-momentum $v^{\pm} = v \cdot n_{\mp} = (v^0 \pm v^{3})/\scriptstyle\sqrt{2}$. Formally, GPDs also depend on the factorization scale, $\mu^2$, typically related to the virtuality of the hard probe $Q^2$. For brevity, the dependence on this variable will be omitted in this work, { \it i.e.} we will consider a fixed scale.
Since in this work we focus on the description of the DVCS amplitude
at the leading order (LO), leading twist (LT) accuracy
we restrict our discussion to quark GPDs only.
For brevity, most of the formulae will be given for a single quark of unspecified flavour, unless superscript ``$\mathrm{DVCS}$'' is used. In such a case, a given quantity should be understood as a sum of contributions coming from the light quarks weighted by the squares of chargers of these quarks.

The polynomiality property states that a given Mellin moment of a GPD is a fixed-order polynomial in $\xi$. For definiteness, we consider the case of quark unpolarized nucleon GPDs:
\begin{eqnarray}
\int_{-1}^1 dx\, x^N H^q(x,\xi,t) &=&
  \sum_{\substack{k=0 \\ \scriptstyle{\rm even}}}^N (2\xi)^k {\cal A}_{N+1,k}(t)
  + \mbox{mod}(N,2)\, (2\xi)^{N+1} {\cal C}_{N+1}(t) \,, \label{eq:polynomiality_H} \\
\int_{-1}^1 dx\, x^N E^q(x,\xi,t) &=&
  \sum_{\substack{k=0 \\ \scriptstyle{\rm even}}}^N (2\xi)^k {\cal B}_{N+1,k}(t)
  - \mbox{mod}(N,2)\, (2\xi)^{N+1} {\cal C}_{N+1}(t) \,, \label{eq:polynomiality_E}
\end{eqnarray}
where
${\cal A}_{n+1,k}(t)$,
${\cal B}_{N+1,k}(t)$
and
${ \cal C}_{N+1}(t)$
are the $t$-dependent coefficients occurring in the form factor decomposition of nucleon matrix elements of the local twist-$2$ operators  \cite{Ji:1998pc,Diehl:2003ny}:
\be
\mathcal{O}_q^{\mu \mu_1 \ldots \mu_N}(0)=
\mathbf{S}
\bar{\psi}(0) \gamma^\mu i
 \overleftrightarrow{D}^{\mu_1}
\ldots
i
 \overleftrightarrow{D}^{\mu_N} \psi(0)\,,
 \label{Def_operator_twist2}
\ee
\be
\begin{aligned}
\left\langle p^{\prime}\left|\mathcal{O}_q^{\mu \mu_1 \ldots \mu_N}(0)\right| p\right\rangle& = \mathbf{S} \bar{u}\left(p^{\prime}\right) \gamma^\mu u(p) \sum_{\substack{k=0 \\
\text { even }}}^N { \cal A}_{N+1, k}^q(t) \Delta^{\mu_1} \ldots \Delta^{\mu_k} P^{\mu_{k+1}} \ldots P^{\mu_N} \\
& +\mathbf{S} \bar{u}\left(p^{\prime}\right) \frac{i \sigma^{\mu \alpha} \Delta_\alpha}{2 m} u(p) \sum_{\substack{k=0 \\
\text { even }}}^N { \cal B}_{N+1, k}^q(t) \Delta^{\mu_1} \ldots \Delta^{\mu_k} P^{\mu_{k+1}} \ldots P^{\mu_N} \\
& +\mathbf{S} \frac{\Delta^\mu}{m} \bar{u}\left(p^{\prime}\right) u(p) \bmod(N, 2) { \cal C}_{N+1}^q(t) \Delta^{\mu_1} \ldots \Delta^{\mu_N},
\end{aligned}
\label{FF_decomp_Diehl}
\ee
where $\mathbf{S}$
denotes symmetrization in all non-contracted Lorentz indices and subtraction of trace terms.

We have chosen skewness $\xi$ to be positive and employ charge-even, $H_+(x,\xi,t)$, and charge-odd, $H_-(x,\xi,t)$, combinations of GPDs defined as
\be
H_\pm(x,\xi,t)=H^q(x,\xi,t)\mp H^q(-x,\xi,t)\,.
\label{Def_Hpm}
\ee
The same definition holds for the GPD $E$, which is also considered in this study. The combinations exhibit the following symmetry:
\be
H_{\pm}(x, \xi, t)=\mp H_{\pm}(-x, \xi, t) \,,
\ee
allowing us to focus on either charge-even or charge-odd GPD component, and to restrict the relevant integration range to $x\in[0;\,1]$, despite the full support domain in this variable is $x\in[-1;\,1]$. In the limit of $\xi=0$, GPDs $H_\pm$ reduce to  $t$-dependent
quark densities, that for $x\ge 0$ gives:
\be
H_\pm (x, \xi=0, t)=  q(x, t) \pm \bar{q}(x, t) \equiv q_{\pm}(x, t) \,.
\label{Def_forward_limit}
\ee
 One may immediately realise that $H_{+}(x, \xi, t)$ only contributes to odd Mellin moments, while $H_{-}(x, \xi, t)$ to even moments. Let us now focus on the interpretation of first few moments:
\bi
\item the zeroth Mellin moment of $H_-(x,\xi,t)$ and $E_-(x,\xi,t)$ give
\be
\int_0^1 d x H_{-} (x, \xi, t)=F_1^q(t) \,, \ \ \ \ \ 
\int_0^1 d x E_{-} (x, \xi, t)=F_2^q(t) \,,
\ee
where $F_{1,2}^q(t)$   are the partonic contributions to the Dirac and Pauli form factors, respectively.
\item
the first Mellin moment of $H_+(x,\xi,t)$ gives
\be
\int_0^1 dx \, x H_+(x,\xi,t)= M_2^q(t)+ \frac{4}{5} d_1(t) \xi^2 \,,
\label{First_Mellin_moment}
\ee
where $M_2^q(t)$ is the $t$-dependent momentum fraction carried by a given quark flavour,
\be
M_2^q(t)= \int_0^1 dx \,  x q_+(x,t) \,,
\ee
and where $d_1(t)$ is the first Gegenbauer coefficient of the $D$-term expansion \cite{Polyakov:1999gs}.
\ei

The elementary leading twist-$2$  leading order elementary amplitudes for  charge-even and  charge-odd GPDs are defined as
\be
\mathcal{H}_{\pm}(\xi, t)=\int_0^1 d x\left[\frac{1}{\xi-x-i \epsilon} \mp \frac{1}{\xi+x-i \epsilon}\right] H_{ \pm}(x, \xi, t)\,,
\label{Def_Elem_Ampl}
\ee
with the imaginary part specified by the GPD values on the cross-over line $x=\xi$:
\be
\im \mathcal{H}_{\pm}(\xi, t)= \pi H_{\pm}(\xi, \xi, t)\,.
\label{Im_part_CFF}
\ee
For brevity, we will refer the  charge-even elementary LO amplitude as the CFF.

\section{Spinless target: case of charge-even GPD $H_{+}$}
\label{Sec_Spinless_case}

In this section, mainly following Ref.~\cite{Muller:2015vha},
we work out the Froissart-Gribov projection of the LO elementary amplitude (\ref{Def_Elem_Ampl})
for the case of the charge-even GPD $H_{+}$ of a spinless hadron. The case of charge-odd GPD
$H_-$, not relevant for DVCS, is presented in Appendix~\ref{App_C_odd}.
We discuss the interpretation of the FG projection in the framework of the dual
parametrization of GPDs and construct a set of sum rules for the
generalized form factors. We argue that these sum rules can be challenged with the
available experimental data for the Compton form factors, forward partonic densities,
and some input from studies of the Mellin moments of GPDs,
\emph{e.g.} calculated
within the lattice QCD framework.

The essence of the  FG projections
\cite{Froissart:1961ux,Gribov:1961ex}
consists in the  reconstruction
of the cross channel partial wave expansion amplitudes from the dispersive
representation of the amplitude in the direct channel.
The derivation of the FG projections for the generalized Compton FF
$\mathcal{H}_{+}$
relies on the once subtracted fixed-$t$ dispersion relation
\cite{Anikin:2007yh,Diehl:2007jb}:
\be
\operatorname{Re} \mathcal{H}_{+}(\xi, t)=\mathcal{P} \int_0^1 d x \frac{2 x H_{+}(x, x, t)}{\xi^2-x^2}+4 D(t)\,,
\label{DispRel_start}
\ee
where $\mathcal{P}$ denotes the principal value integration and the subtraction constant $D(t)$ is the so-called $D$-term form factor.
The latter is given by the sum of coefficients coming from the Gegenbauer expansion of the $D$-term
\cite{Polyakov:1999gs}.

The expansion of the Compton FF in the cross channel SO$(3)$ partial waves
naturally arises once considering the $t$-channel counterpart of the DVCS reaction:
\be
\gamma^*(q)+\gamma\left(-q^{\prime}\right) \rightarrow h\left(p^{\prime}\right)+\bar{h}(-p)\,,
\label{Reaction_Cross_channel}
\ee
with the Mandelstam variable $t=(q-q')^2$ playing the role of
the invariant CMS energy and the $t$-channel scattering angle $\theta_t$ defined
as the angle between $\vec{q}$ and $\vec{p}'$ in the
$\gamma^*(q)\, \gamma(-q')$-center-of-mass frame.
For the Compton FF $\mathcal{H}_{+}$ the PW expansion involves only even spin PWs and takes the following
form:
\be
\mathcal{H}_{+}\left(\cos \theta_t, t\right)=
\sum_{\substack{J=0 \\ \text { even }}}^\infty
F_J(t) P_J(\cos \theta_t)\,,
\label{Def_PW_expansion_cross_channel}
\ee
where $P_J(\cos \theta_t)$ stand for the Legendre polynomials
and the PW expansion coefficients $F_J(t)$ are defined as
\footnote{The relation between the PWs
$a_J(t)$ defined in  Refs.~\cite{Muller:2014wxa,Muller:2015vha} and the present
is $F_J(t) \equiv (2J+1) \,a_J(t).$}
\be
F_J(t) = \frac{2J+1}{2} \int_{-1}^1 d\left(\cos \theta_t\right) P_J\left(\cos \theta_t\right) \mathcal{H}_{+}\left(\cos \theta_t, t\right).
\label{Def_FJ_FF}
\ee

After crossing the reaction (\ref{Reaction_Cross_channel}) back to the direct channel
\be
\gamma^*(q)+ {h}(p) \rightarrow  \gamma\left(q^{\prime}\right)+ h\left(p^{\prime}\right),
\label{Reaction_Direct_channel}
\ee
within the usual DVCS kinematics (large-$Q^2$ and $s=(p+q)^2$, fixed $x_B \equiv \frac{Q^2}{2 p \cdot q}$, $-t$ of hadronic mass scale $m^2$), the analytically continued expression for the cosine of the $t$-channel scattering angle $\theta_t$,
 up to power suppressed corrections, becomes
\be
\cos \theta_t \rightarrow-\frac{1}{\xi \beta}+\mathcal{O}\left(1 / Q^2\right),
\label{Cos_theta_approx}
\ee
where
\be
\beta=\sqrt{1-\frac{4 m^2}{t}}
\label{Def_beta}
\ee
is the usual relativistic velocity factor.

In the present analysis we neglect effects related to the target mass and set $\beta=1$. The discussion describing the complications associates with $\beta \ne 1$ is provided at the end of the Section.
To establish the FG projection formula for the
form factors (\ref{Def_FJ_FF}) we rely on
the dispersion relation (\ref{DispRel_start}) analytically continued to the $t$-channel:
\be
\mathcal{H}_{+}\left(\cos \theta_t, t\right) =\int_0^1 d x \frac{2 x \cos ^2 \theta_t}{1-x^2 \cos ^2 \theta_t} H_{+}(x, x, t)+4 D(t)\,.
\label{Disp_rel_t_channel}
\ee
We would like to stress that no regulator is required for the singularity in the denominator, as for the physical domain of the $t$-channel, where $|\cos \theta_t| \le 1$, the pole in $x$
remains outside of the domain of integration.
We also suppose $H_{+}(x, x, t)$ vanishes fast enough for $x=1$ to keep the integral regular.

We make use of the Neumann integral representation for the Legendre functions
of the second kind,
$\mathcal{Q}_J$,
with  integer
$J \ge 0$
(see {\it e.g.} Ref.~\cite{Yanke1}):
\be
\frac{1}{2} \int_{-1}^1 d z' P_J(z') \frac{1}{z-z'}=\mathcal{Q}_J\left(z\right)\,.
\label{Neumann_representation}
\ee
Note that $\mathcal{Q}_J(z)$ are defined in the complex-$z$ plane with the cut along the segment $[-1;\,1]$. The
somewhat more familiar second kind Legendre functions of a real argument $x \in (-1;\,1)$, ${\rm Q}_J(x)$, are defined as the following discontinuity\footnote{In particular, the default realization in the Wolfram Mathematica \cite{Wmath} deals with these latter second kind Legendre functions.}:
\be
{ \rm Q}_J(x)=\frac{1}{2}\left({ \cal Q}_J(x+ i0)+{ \cal Q}_J(x- i0)\right)\,.
\label{Def_conventional_Legendre_second_kind}
\ee

After plugging the dispersion relation representation of the Compton FFs~\eqref{Disp_rel_t_channel} and the Neumann integral representation~\eqref{Neumann_representation}
into the definition of the SO$(3)$ PWs~\eqref{Def_FJ_FF}, and interchanging the order of integration we get:
\begin{itemize}
\item for $J=0$:
\be
F_{J=0}(t)=2 \int_0^1 d x\left(\frac{\mathcal{Q}_0(1 / x)}{x^2}-\frac{1}{x}\right) H_{+}(x, x, t)+4 D(t)\,.
\label{FJ0}
\ee
\item for even positive $J$:
\be
F_{J>0}(t) =2 (2J+1)\int_0^1 d x \frac{\mathcal{Q}_J(1 / x)}{x^2} H_{+}(x, x, t)\,,
\label{F_Jne0}
\ee
\end{itemize}

Since for small $x$
\be
\frac{\mathcal{Q}_J(1 / x)}{x^2} = \frac{J!}{(2J+1)!!} x^{J-1} + {\cal O }(x^{J+1})\,,
\label{Small_x_for_Qs}
\ee
the integrals in Eqs.~\eqref{F_Jne0} and~\eqref{FJ0} are well convergent under the usual Regge phenomenology assumptions for small-$x$ asymptotic behavior of GPDs
on the cross over line:
\be
H_{+}(x, x, t) \sim x^{-\alpha(t)} \ \  \text{with}  \ \ \alpha(0)<2\,.
\ee

\begin{figure}[H]
 \begin{center}
\includegraphics[width=0.4\textwidth]{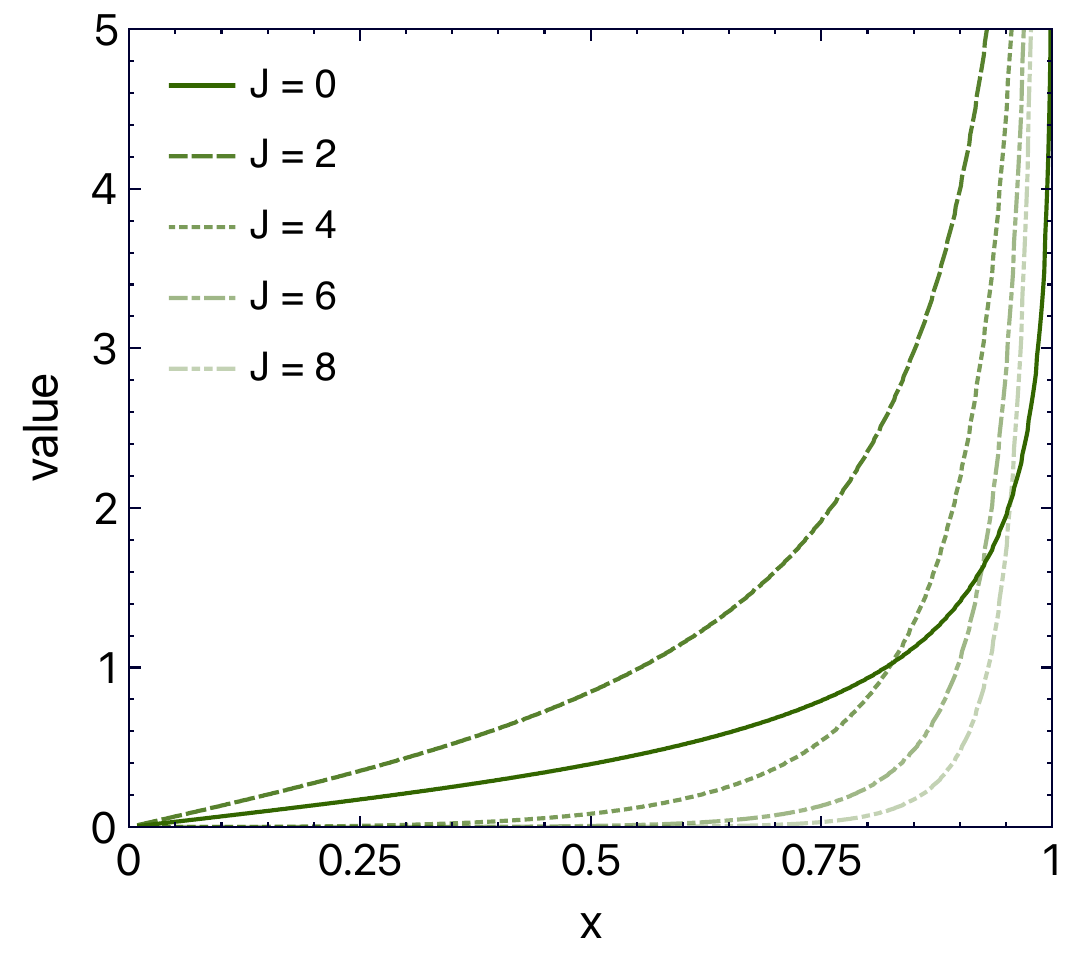}
\caption{First weight functions $2(2J+1) \left(\frac{Q_J(1/x)}{x^2}-\delta_{J0} \frac{1}{x} \right)$ entering definition of the $F_J(t)$ projections.}
 \label{Fig_Q_functions}
\end{center}
\end{figure}

In Fig.~\ref{Fig_Q_functions} we show several first weight functions $2(2J+1) \left(\frac{Q_J(1/x)}{x^2}-\delta_{J0} \frac{1}{x} \right)$ entering definition of the FG projections
(\ref{FJ0}),
(\ref{F_Jne0}). It is remarkable that for larger $J$ the small-$x$ region is progressively damped, as follows from the
asymptotic expansion (\ref{Small_x_for_Qs}). This makes the high-$J$ FG projections
more sensitive to the large-$x_B$ behavior of the Compton FFs. Therefore, the FG projections can be
useful to study the interplay of $\xi$- and $t$-dependencies of the Compton FFs and the buildup of the skewness effect for large $\xi$.

In a rigorous sense, the interpretation of the form factors $F_J(t)$ in terms of definite angular momentum of hadron-antihadron pair $h(-p) \, \bar{h}(p')$
in the reaction (\ref{Reaction_Cross_channel}), in the spirit of the analysis of Ref.~\cite{Polyakov:1998ze}, relies on the highly non-trivial issue of analytic continuation in $t$ from the spacelike to timelike region. An attempt to
perform this type of analytic continuation in a constructive manner was presented in the dispersive analysis of Ref.~\cite{Pasquini:2014vua} for the case of the  $D$-term form factor.

An alternative way of interpreting of the form factors $F_J(t)$, which does not necessarily imply the analytic continuation in $t$ to the cross channel, arises in the framework
of the  dual parametrization  of GPDs
\cite{Polyakov:2002wz,Muller:2014wxa}.
In this approach GPDs are presented as double partial wave expansions (both in the conformal and cross channel SO$(3)$ partial waves). The use of the conformal basis ensures
the diagonalization of the LO evolution operator, while the cross channel SO$(3)$ PW expansion results in a factorization of $x$, $\xi$ and $t$ dependencies of a given GPD.

For the $C=+ 1$ quark%
\footnote{A generalization for the gluon GPD case was presented in Ref.~\cite{Semenov-Tian-Shansky:2010vsl}.}
GPD
$H_{+}$
this expansion takes the form of
\be
H_{+}(x, \xi, t)=2 \sum_{\substack{n=1 \\ \text { odd  }}}^{\infty} \sum_{\substack{l=0 \\ \text { even }}}^{n+1} B_{n, l}(t) \, \theta \left(1-\frac{x^2}{\xi^2} \right)\left(1-\frac{x^2}{\xi^2}\right) C_n^{3 / 2}\left(\frac{x}{\xi}\right) P_l\left(\frac{1}{\xi}\right),
\label{Dual_Param_Cplus}
\ee
where $n+2$ corresponds to the conformal spin and $l$ refers to the cross-channel angular momentum.
The corresponding double partial wave amplitudes (generalized form factors), $B_{n, l}(t)$, for odd
$n$
are generated through the Mellin transform of a set of the charge-even forward-like functions
$Q_{2 \nu}(y,t)$ of an auxiliary variable $y$:
\be
B_{n, n+1-2 \nu}(t)=\int_0^1 d y y^n Q_{2 \nu}(y, t)\,, \quad\text{or equivalently,}\quad B_{n, l}(t)=\int_0^1 d y y^n Q_{n+1-l}(y, t)\,.
\label{Def_Bnl}
\ee
To the leading logarithmic accuracy, the flavor non-singlet  coefficients
(\ref{Def_Bnl}) are renormalized multiplicatively with the same anomalous dimensions
as meson distribution amplitudes, while the singlet case is complicated by mixing under evolution with gluons, see discussion in Sect.~3.7.2 of Ref.~\cite{Diehl:2003ny}.

The crucial role in the double PW expansion approach is played by the parameter
$\nu=0,\,1,\,2, \ldots$ defined by the difference between the conformal spin $n+2$
and the cross channel angular momentum $l$:
\be
n+2-l \equiv 2 \nu +1 \,,
\label{Def_nu_parameter}
\ee
providing a classification of PW contributions into GPDs.
The function
 $Q_0(y,t)$%
\footnote{Not to be mixed up with the second kind Legendre function defined in Eq.~\eqref{Def_conventional_Legendre_second_kind}.} in (\ref{F_J=0_through_N}) is fixed in terms of $t$-dependent PDFs:
\be
Q_0(y,t)=q_{+}(y, t)-\frac{y}{2} \int_y^1 \frac{d z}{z^2} q_{+}(z, t)\,,
\label{Def_Q0}
\ee
in accordance with the forward limit constraint (\ref{Def_forward_limit}), while
the functions $Q_{2 \nu}(y,t)$ with $\nu>0$ contain the truly non-forward information
encoded in GPDs.

For odd $N \ge 1$ the Mellin moments of the GPD
(\ref{Dual_Param_Cplus}) read:
\be
\int_0^1 d x x^N H_{+}(x, \xi, t) =\sum_{\substack{k=0 \\ \text { even }}}^{N+1} h_{N, k}(t) \xi^k,
\label{Mellin_moments_Hplus}
\ee
where the coefficients at powers of $\xi$ are expressed as
\be
\begin{aligned}
h_{N, k}(t) & = \sum_{\substack{n=1 \\
\text { odd }}}^N \sum_{\substack{l=0 \\
\text { even }}}^{n+1} B_{n,l}(t)(-1)^{\frac{k+l-N-1}{2}} \frac{\Gamma\left(1-\frac{k-l-N}{2}\right)}{2^k \Gamma\left(\frac{1}{2}+\frac{k+l-N}{2}\right) \Gamma(2-k+N)} \frac{(n+1)(n+2) \Gamma(N+1)}{\Gamma\left(1+\frac{N-n}{2}\right) \Gamma\left(\frac{5}{2}+\frac{N+n}{2}\right)}\,.
\end{aligned}
\label{Mellin_moments_coefs_through_Bnl}
\ee
Particularly, for the first Mellin moment
(\ref{First_Mellin_moment})
providing an access to hadronic
matrix elements of the quark QCD EMT, we get:
\be
 \int_0^1 d x x \, H_{+}(x, \xi, t)= \frac{6 B_{1,2}(t)}{5}+  \xi ^2 \left(\frac{4 B_{1,0}(t)}{5}-\frac{2 B_{1,2}(t)}{5}\right)\,.
\ee

The double partial wave expansion
(\ref{Dual_Param_Cplus})
has to be understood as an ill-defined sum of generalized
functions: although each individual term of this expansion has the central region support  $|x| \le \xi$, it does not imply that the resulting GPD vanishes outside this region.
This expansion  is provided a rigorous meaning with help of a resummation based on the
Shuvaev-Noritzch transform
\cite{Noritzsch:2000pr}
resulting in the following integral representations:
\be
H_{+}(x, \xi, t)=\sum_{\nu=0}^{\infty} \int_0^1 d y\left[K_{2 \nu}(x, \xi \mid y) - K_{2 \nu}(-x, \xi \mid y)\right] y^{2 \nu} Q_{2 \nu}(y, t)\,,
\label{Hplus_through_Qs}
\ee
with the convolution kernels
$K_{2 \nu}(x, \xi \mid y)$ non-vanishing for $-\xi \le x \le 1$ and
expressed in terms of the elliptic integrals \cite{Polyakov:2002wz,Polyakov:2009xir}.
The dual parametrization of GPDs was found
to be completely equivalent to the
GPD representation based on the Mellin-Barnes integral approach \cite{Mueller:2005ed,Kumericki:2007sa}. The set of formulae connecting these two representations has been
elaborated in Ref.~\cite{Muller:2014wxa}.
In Appendix~\ref{App_conformal_moments} we present
a brief summary of formulae expressing the double PW expansion coefficients within the two representations.

For phenomenological applications of the dual parametrization
of GPDs, as well as of the Mellin-Barnes integral approach framework, it is
essential to justify a truncation of $\nu$-summation
in (\ref{Hplus_through_Qs})
accounting for
a finite number of PWs
\bi
\item
In the dual parametrization the $\nu=0$ contribution (\ref{Def_Q0}) entirely defined
in terms of $q_+(y,t)$ corresponds to the so-called  ``minimalist dual model''.
However, as pointed out in \cite{Kumericki:2008di,Guzey:2008ys}, the  minimalist model can not properly describe the H1 data  for small-$x_B$ \cite{H1:2007vrx},
as it produces an inconsistent skewness ratio
$
\frac{H_{+}(x, x, t=0)}{H_{+}( x, 0, t=0)}
$
for $x \sim 0$.
\item In order to control the  skewness effect for small-$x$ and the evolution effects in a broad range of $Q^2$,
it turns out necessary to account for at least three contributions corresponding to $\nu=0,\,1,\,2$. Within the Mellin-Barnes
framework this yields  a basis for the
Kumeri\v{c}ki-M\"{u}ller (KM) phenomenological model \cite{Kumericki:2015lhb}
providing a satisfactory description of the DVCS data at small-$x_B$.
\ei

In the dual parametrization framework, the FG projection form factors  (\ref{FJ0}), (\ref{F_Jne0})
can be expressed
through the $J-1$-th Mellin moments of the so-called GPD quintessence function
$N(y,t) \equiv \sum_{\nu=0}^\infty y^{2 \nu} Q_{2 \nu}(y,t)$
\cite{Polyakov:2007rw,Polyakov:2007rv}:
\be
F_{J=0}(t)
=4 \operatorname{Reg} \int_0^1 \frac{d y}{y}\left(N(y, t)-Q_0(y, t)\right);
\label{F_J=0_through_N}
\ee
and
\be
F_{J>0}(t)
=
4\int_0^1 d y y^{J-1} N(y, t)\,.
\label{FJ_through_N}
\ee
The GPD quintessence function $N(y,t)$ occurring in Eqs.~(\ref{F_J=0_through_N}) and~(\ref{FJ_through_N})
can be recovered from the absorptive part of LO
elementary amplitude   (\ref{Im_part_CFF})
by the inverse Abel tomography procedure \cite{Polyakov:2007rv,Moiseeva:2008qd}:
\be
\begin{aligned}
N(y, t) & = \frac{1}{2 \pi} \frac{\sqrt{2 y}(1+y)}{\sqrt{1+y^2}} \operatorname{Im} {\cal H}_+\left(\frac{2 y}{1+y^2}, t\right) \\
& -\frac{1}{2 \pi} \frac{y\left(1-y^2\right)}{\left(1+y^2\right)^{\frac{3}{2}}} \int_{\frac{2 y}{1+y^2}}^1 d x \frac{1}{\left(x-\frac{2 y}{1+y^2}\right)^{\frac{3}{2}}}\left(\frac{1}{\sqrt{x}} \operatorname{Im} {\cal H}_+(x, t)-\sqrt{\frac{1+y^2}{2 y}} \operatorname{Im} {\cal H}_+\left(\frac{2 y}{1+y^2}, t\right)\right)\,.
\end{aligned}
\label{N_function_Abel}
\ee
The integral in (\ref{F_J=0_through_N}) may require some form of regularization of a singularity at $y=0$ inherited from the Regge-like behavior of the partonic densities. This leads to a possible manifestation of
the $J=0$ fixed pole contribution into the $D$-term form factor, see discussion in \cite{Muller:2015vha}.

The equivalence of the definitions (\ref{F_J=0_through_N}), (\ref{FJ_through_N}), and, respectively, (\ref{FJ0}), (\ref{F_Jne0}) was established in
\cite{Muller:2014wxa}
by computing the $(J-1)$-th Mellin moment of (\ref{N_function_Abel}) as
\be
\int_0^1 d y y^{J-1} N(y, t)=\int_0^1 d x\left[\frac{1}{\sqrt{x}} \frac{d}{d x} R_J(x)\right] H_{+}(x, x, t)\,,
\ee
where the  auxiliary functions $R_J(x)$ can be expressed  through the Legendre functions of the second kind (\ref{Neumann_representation}):
\be
\frac{1}{\sqrt{x}} \frac{d}{d x} R_J(x)=\frac{1}{\sqrt{x}} \frac{d}{d x} \int_0^x \frac{d w}{\sqrt{2} \sqrt{w}}\left(\frac{1-\sqrt{1-w^2}}{w}\right)^{J+\frac{1}{2}} \frac{1}{\sqrt{x-w}}=\frac{\frac{1}{2}+J}{2} \frac{2 \mathcal{Q}_J(1 / x)}{x^2}\,.
\ee

The physical contents of the form factors (\ref{F_J=0_through_N}), (\ref{FJ_through_N}) can now be revealed in a particularly simple manner.
For even $J$, the $J-1$-th Mellin moment of the GPD quintessence function (\ref{N_function_Abel}) corresponds to the sum of generalized form factors with the cross channel  angular momentum label $l=J$:
\be
F_{J=0}(t) =4 \sum_{\substack{n=1 \\ \text { odd }}}^{\infty} B_{n, 0}(t)=4\sum_{\nu=1}^\infty B_{2 \nu-1, 0} (t)\,,
\label{F0_dual_parametrization}
\ee
and
\be
F_{J>0}(t) =4 \sum_{\substack{n=J-1 \\ \text { odd }}}^{\infty} B_{n, J}(t) =4\sum_{\nu=0}^\infty B_{J+2 \nu-1, J} (t)\,.
\label{FJ_dual_parametrization}
\ee
Truncating the summation in $\nu$
makes it possible to turn
Eqs. (\ref{F0_dual_parametrization}), (\ref{FJ_dual_parametrization})
into a set of constructive sum rules quantifying the hadron target
response on the spin-$J$ excitation induced by the QCD string operators
(\ref{QCD_String_Operators}).

Indeed, according to Eq.
(\ref{Mellin_moments_coefs_through_Bnl}),
the generalized form factors
$B_{n,l}(t)$
are in a direct relation to the coefficients
$h_{N,k}(t)$
at powers of
$\xi$
of the Mellin moments of the GPD (\ref{Mellin_moments_Hplus}). In turn,
according to the standard exposition of the GPD polynomiality property,
the coefficients
$h_{N,k}(t)$ correspond to form factors occurring in the decomposition
of hadronic matrix elements of local quark twist-$2$ operators, that
can be studied with the methods of lattice QCD, see {\it e.g.} Refs.~\cite{Hagler:2003jd,Hagler:2009ni}.

For example, truncating at $\nu=1$ (``next-to-minimalist'' contribution)
for the $J=0$ FG projection we get:
\be
&&
F_{J=0}(t)=4 \left( B_{1,0}(t)+ \ldots \right) =
\frac{5}{3} h_{1,0}(t)+5 h_{1,2}(t) +\genfrac\{\}{0pt}{0}{\text{contribution of conformal PWs}}{\text{with} \; \nu \ge 2}\,,
\label{FJ=0_SR}
\ee
and for
$J=2$
\begin{align}
F_{J=2}(t) = 4 \left( B_{1,2}(t)+B_{3,2}(t) + \ldots \right) = -\frac{7}{6} h_{1,0}(t)+9 h_{3,0}(t)+\frac{21}{2} h_{3,2}(t)
+\genfrac\{\}{0pt}{0}{\text{contribution of conformal PWs}}{\text{with} \; \nu \ge 2}\,,
\label{FJ=2_SR}
\end{align}
where the coefficients $h_{1,0}(t)$  and $h_{3,0}(t)$ are fixed
from the Mellin moments of $t$-dependent PDFs:
\be
&&
h_{1,0}(t)= \frac{6}{5} B_{1,2}(t)= \int_0^1 dx x q_+(x,t) \equiv M_2(t)\,, \nn \\ &&
h_{3,0}(t)= \frac{10}{9} B_{3,4}(t)= \int_0^1 dx x^3 q_+(x,t)\,,
\ee
and
$h_{1,2}(t) \equiv  \frac{4}{5} d_1(t)$\,.

A generalization of the sum rules (\ref{FJ=0_SR}), (\ref{FJ=2_SR})
with truncation at arbitrary $\nu$ (as well as for arbitrary $J$), is straightforward. At any order it
requires  knowledge of a finite number of the Mellin moments of GPDs, though the resulting
expressions may be somewhat bulky.
The further strategy can be twofold.
\bi
\item  The FG projection form factors $F_{J}(t)$ represent observable quantities that can be directly (without a need for deconvoluting GPDs) extracted from DVCS data in analyses like \cite{Kumericki:2007sa,Moutarde:2018kwr,Moutarde:2019tqa,Guo:2022upw, Guo:2023ahv}. On the other hand, the right-hand side of the sum rules can be computed within
existing phenomenological GPD model.
This provides a discrimination between models and allows to  study the impact of higher-$\nu$ conformal PWs.
\item
Provided the development of lattice methods for
computing the Mellin moments of GPDs, see {\it e.g.} \cite{Hagler:2003jd,Hagler:2009ni,Bhattacharya:2023ays},
the sum rules like
(\ref{FJ=0_SR}), (\ref{FJ=2_SR}) can be directly used to test lattice predictions against experimental data and phenomenological GPD models.

\ei

In our consideration we, so far, neglected the target mass, so that hadron helicities were truly conserved quantum numbers, that excluded mixing among the SO$(3)$ PWs. Let us now discuss the complications
associated with non-zero target mass.
A detailed treatment of this issue
represents a formidable task and may require the use of the refined theoretical methods developed {\it e.g.} in Refs.~\cite{Braun:2011zr,Braun:2011dg,Braun:2012bg}.

Firstly, it is fair to recognize that the derivation of the FG projections along the lines
of Eqs.~(\ref{Disp_rel_t_channel})-(\ref{F_Jne0}) looks somewhat problematic for $\beta \ne 1$.
Indeed, after restoring the $\beta$ factor, the  dispersion relation (\ref{DispRel_start}) analytically continued to the $t$-channel reads
\be
 \mathcal{H}_{+}\left(\cos \theta_t, t\right) =  \int_0^1 d x \frac{2 x  \beta^2 \cos ^2 \theta_t}{1-x^2 \beta^2  \cos ^2 \theta_t} H_{+}(x, x, t)+4 D(t).
\label{Disp_rel_t_channel_beta}
\ee
Note that, as far as we stay in the physical region of the $t$-channel
with $t> 4 m^2$ ($0 < \beta < 1$), $|\cos \theta_t| \le 1$,
no regulator is required in the denominator of the integrand of (\ref{Disp_rel_t_channel_beta}).
However, the calculation of the FG projections
from the representation  (\ref{Disp_rel_t_channel_beta}) employing the Neumann integral
(\ref{Neumann_representation})
results in appearing of
the $1/\beta$ factor in the argument
of the corresponding associated Legendre functions. This makes the generalization of Eq.~(\ref{F_Jne0})
to look like
\be
F_{J>0}(t) =2 (2J+1)\int_0^1 d x \frac{\mathcal{Q}_J \left( \frac{1}{x \beta}\right)}{ \beta x^2}
H_{+}(x, x, t)\,.
\label{F_Jne0_beta}
\ee
We see that the analytic continuation of Eq.~(\ref{F_Jne0_beta})
to the values of $t$ corresponding to the DVCS channel is troublesome,
as for $\beta>1$ and
$x > \frac{1}{\beta}$
the second kind Legendre function,
$\mathcal{Q}_J \left( \frac{1}{x \beta}\right)$,
possesses a cut inside the integration domain.
A way out could be restricting the upper limit of the integration
to $x_{\max}= \frac{1}{\beta}$. This might look appealing also from the perspective
of excluding the contribution of the nonphysical domain $x_B > x_{B \,, \max} \equiv \frac{2}{1+ \beta}$ out of the scope of dispersive analysis.
Unfortunately, to the best of our knowledge, the reduction of the integration domain in the dispersion relation
(\ref{DispRel_start}) can not be justified.
Derivation of the dispersion relation
(\ref{DispRel_start})
departing from the once-subtracted fixed-$t$ dispersion relation
in the energy variable $\frac{s-u}{4 m}$ for the Compton amplitude
implies, at the final step, the strict use of the generalized Bjorken limit, see {\it e.g.} Ref.~\cite{Muller:2015vha}. This does not allow to properly implement the threshold corrections and ultimately sets
the upper $x$-integration limit to $1$ in the dispersion relation (\ref{Disp_rel_t_channel_beta}).

In order to circumvent the aforementioned difficulties related to a direct inclusion of $\beta \neq 1$ in the dispersion relation, we turn to the framework of the dual parametrization. We consider a modified version of the double PW expansion for the $C=+1$ GPD, \emph{cf.} Eq.~\eqref{Dual_Param_Cplus}. The modification is the explicit inclusion of
threshold corrections in the  summation of the $t$-channel spin-$l$ exchanges:
\be
H_{+}(x, \xi, t)=2 \sum_{\substack{n=1 \\ \text { odd  }}}^{\infty} \sum_{\substack{l=0 \\ \text { even }}}^{n+1}  \beta^l {\bar{B}}_{n, l}(t) \, \theta \left(1-\frac{x^2}{\xi^2} \right)\left(1-\frac{x^2}{\xi^2}\right) C_n^{3 / 2}\left(\frac{x}{\xi}\right) P_l\left(\frac{1}{\xi \beta}\right).
\label{Dual_Param_Cplus_beta}
\ee
The expansion is performed in $P_l\left(\frac{1}{\xi \beta}\right) = P_l(\cos \theta_t)$, rather than  $P_l\left(\frac{1}{\xi  }\right)  = P_l(\beta \cos \theta_t)$
\footnote{We employ  (\ref{Cos_theta_approx}) to the leading order accuracy, and, since $l$ is even, drop the $(-1)$ factor in the argument of the Legendre polynomials.}. In addition, the $l$-th SO$(3)$ PW now includes the factor $\beta^l$, making the resulting GPD look singular for $t \to 0$. We note, that in the cross channel the $\beta^l$ factor corresponds to the usual suppression of the $l$-th PW.
Similarly to Eqs.~(\ref{Def_Bnl}) and~\eqref{Mellin_moments_coefs_through_Bnl}, we introduce a set of the forward-like functions, $\bar{Q}_{2 \nu}(y,t)$, generating the form factors
$\bar{B}_{n,l}(t)$:
\be
\bar{B}_{n, n+1-2 \nu}(t)=\int_0^1 d y y^n \bar{Q}_{2 \nu}(y, t)\,, \quad\text{or equivalently,}\quad {\bar B}_{n, l}(t)=\int_0^1 d y y^n \bar{Q}_{n+1-l}(y, t)\,,
\label{Def_Bnl_beta}
\ee
and we provide the expression for
the coefficients at powers of $\xi$ of the $N$-th Mellin moment
(\ref{Mellin_moments_Hplus}):
\be
\begin{aligned}
h_{N, k}(t)=\sum_{\substack{n=1 \\
\text { odd }}}^N \sum_{\substack{l=0 \\
\text { even }}}^{n+1} \beta^{l+k-N-1} \bar{B}_{n,l}(t)(-1)^{\frac{k+l-N-1}{2}} \frac{\Gamma\left(1-\frac{k-l-N}{2}\right)}{2^k \Gamma\left(\frac{1}{2}+\frac{k+l-N}{2}\right) \Gamma(2-k+N)} \frac{(n+1)(n+2) \Gamma(N+1)}{\Gamma\left(1+\frac{N-n}{2}\right) \Gamma\left(\frac{5}{2}+\frac{N+n}{2}\right)}\,.
\end{aligned}
\label{Mellin_moments_coefs_through_Bnl_beta}
\ee

We now would like to establish a link
between the FG projections
(\ref{F0_dual_parametrization}),
(\ref{FJ_dual_parametrization})
constructed by means of the Abel tomography procedure
(\ref{N_function_Abel})
in terms of the generalized FFs $B_{n,l}$ of the double PW expansion
(\ref{Dual_Param_Cplus}),
and the set of the generalized FFs $\bar{B}_{n,l}$
occurring in the improved double PW expansion
(\ref{Dual_Param_Cplus_beta}),
accounting for the
threshold corrections.
This requires working out a relation between the generalized FFs
$B_{n,l}$
and
${\bar{B}}_{n, l}$, which
correspond  to a definite value of the cross channel angular momentum $J=l$.

In order to find a relation between these two quantities, one has to re-expand
(\ref{Dual_Param_Cplus_beta})
over the basis formed by the Legendre polynomials
$P_l( \frac{1}{\xi}) \equiv P_l( \beta  \cos \theta_t)$.
A similar type of series transformation was addressed in Refs.~\cite{Polyakov:1998ze,Diehl:2000uv}
for the case of two-hadron Generalized Distribution Amplitudes (GDAs).
In our case, it can be done by an iterative procedure re-expressing the set of the generalized FFs $B_{n,l}$ through  $\bar{B}_{n,l}$ by
matching the two double PW expansions.
Such iterative procedure goes as follows:
\bi
\item We require the expansions (\ref{Dual_Param_Cplus_beta}) and (\ref{Dual_Param_Cplus}) to give the same
coefficients $h_{N,k}(t)$ at power $\xi^k$ of $N$-th Mellin moments.

\item
The coefficients $h_{N,k}(t)$ must be regular in the $t \to 0$ limit. In the case of expansion
(\ref{Dual_Param_Cplus_beta}),
this requires to assume specific singularities for the generalized FFs
${\bar{B}}_{n, l}(t)$ at $t=0$.

\item
For any odd $N \ge 1$ the expansion
(\ref{Dual_Param_Cplus_beta}) results in the regular coefficient $h_{N,0}(t)$. Therefore, we conclude that
$\bar{B}_{n,n+1}(t)={B}_{n,n+1}(t)$. This also means that the $\nu=0$ forward-like function does not require any modification, \emph{i.e.}
$\bar{Q}_{0}(y,t)= {Q}_{0}(y,t)$, and it remains to be exclusively  fixed by the $t$-dependent PDF, as in Eq.~(\ref{Def_Q0}).
\item
Now we turn to the coefficient $h_{N,2}(t)$. It obtains contributions from the  forward-like functions with $\nu=0$ and $\nu=1$. To keep it regular for $t=0$ one has to assume
\be
B_{n, n-1}(t)=\bar{B}_{n, n-1}(t)-\left(1-\beta^2\right)  \left( \frac{1}{2}-n\right) \bar{B}_{n, n+1}(t).
\label{Bnnm1_modification}
\ee
Thus, the generalized FF $B_{n, l=n-1}(t)$
obtains through mixing a contribution from $l+2$-th PW.
The mixing (\ref{Bnnm1_modification}) is equivalent to the following relation between the $\nu=1$ forward-like functions:
\be
{Q}_2(y, t)=\bar{Q}_2 (y, t) -\left(1-\beta^2\right)\left(\frac{3}{2} \bar{Q}_0(y, t)+y \bar{Q}_0^{\prime}(y, t)\right).
\label{Mixing_Q2}
\ee

\item Similarly, by considering the coefficient $h_{N,2}(t)$, which
obtains contributions from the forward-like functions with
$\nu=0,\,1,\,2$, we conclude that the generalized FF $B_{l+3, l}(t)$
obtains through mixing a contribution from $l+2$-th and $l+4$-th PWs; and the $Q_4(y,t)$ can be expressed in terms of
$\bar{Q}_4(y,t)$,  $\bar{Q}_2(y,t)$ and  $\bar{Q}_0(y,t)$.

\item  The process can be further iterated. The mixing for
$B_{l+2 \nu -1, l}(t)$ involves higher spin contributions up to $l+2\nu$.
\ei

From this discussion we conclude, that  $J$-th FG projections expressed through Eqs. (\ref{F0_dual_parametrization}), (\ref{FJ_dual_parametrization}) obtain an
admixture of higher spin contribution involving a number of cross channel SO$(3)$ PWs increasing with the index $\nu$ (\ref{Def_nu_parameter}).
The quantitative effect of this mixing is not that easy to estimate, as it strongly depends on the number of PWs in $\nu$
one has to take into account in a satisfactory GPD model. For example, the ``minimalist dual model'' accounting only for $\nu=0$ contribution does not produce any mixing; and the corresponding $J$-th FG projections exactly quantify target's response on the  the spin-$J$ cross channel excitation. For a GPD model including a limited number PWs in $\nu$
(like the sea quark part of the KM model \cite{Kumericki:2015lhb} accounting for the $\nu=0,\,1,\,2$ PWs)
the mixing can be easily resolved with help of equations like (\ref{Mixing_Q2}). However, the numerical relevance of this mixing, and its dependence on the value of $J$ of the FG projection deserves a separate study and will be addressed elsewhere.
We conclude, that a necessity to include a large number of PWs in $\nu$ may lead
to a large admixture of contributions of higher spins exchanges into a given
FG projection and this may barely destroy the interpretation
in terms of target's response to the spin-$J$
cross channel excitation.

\section{Froissart-Gribov projection for the spin-$\nicefrac{1}{2}$ target}
\label{Sec_Nucleon_target}

Now we turn to the case of spin-$\nicefrac{1}{2}$ target, which, for brevity, we will refer to as a nucleon.
According to  Sect.~4.2 of Ref.~\cite{Diehl:2003ny}, for such target the appropriate combinations of GPD to perform the expansion in the cross-channel SO$(3)$ partial waves are:
\be
&&
H^{(E)}_{\pm}(x, \cos \theta_t,t) =H_{\pm}(x, \cos \theta_t,t)+
  \tau
E_{\pm}(x, \cos \theta_t,t)\,,
\\ &&
H^{(M)}_{\pm}(x, \cos \theta_t,t)=H_{\pm}(x, \cos \theta_t,t)+ E_{\pm}(x, \cos \theta_t,t)\,,  \ \ \ \
\ee
where $\tau \equiv t/(4m^2)$.
These are the electric and magnetic combinations, respectively, here, separately defined for charge parity $C=\pm 1$ components of GPDs. From the cross-channel perspective
(\ref{Reaction_Cross_channel}), $H^{(E)}_{\pm}(x, \cos \theta_t,t)$ corresponds to the
helicities of nucleon and antinucleon couple to $|\lambda-\lambda'|=0$, while $H^{(M)}_{\pm}(x, \cos \theta_t,t)$ to $|\lambda-\lambda'|=1$.
Furthermore,
\bi
\item
the electric nucleon GPD has to be expanded in the rotation functions
(see {\it e.g} App. A.2 of \cite{Martin:102663})
$d_{00}^J(\cos \theta_t) \equiv P_J(\cos \theta_t)$,
analogously to the case of spinless target GPD;
\item the magnetic nucleon GPD has to be expanded in the rotation functions
$d_{01}^J(\cos \theta_t) \equiv (J(J+1))^{-1/2} \sin \theta_t P'_J(\cos \theta_t)$,
where $ P'_J(z)$ stands for the derivative of the $J$-th Legendre polynomial.
\ei

The electric combinations $H_{\pm}^{(E)}(x,\xi,t)$  are normalized  according to:
\be
&&
\int_0^1 dx \, x H_+^{(E)}(x,\xi,t)= M_2^q(t)\left(1- \tau \right)+ \tau 2 J^q(t)+ \frac{4}{5} \left(1- \tau \right)d_1(t) \xi^2 \,, \\ &&
\int_0^1 dx \,   H_-^{(E)}(x,\xi,t)= G^{(E) \, q}(t)\,.
\ee
The normalisation of the magnetic combinations $H^{(M)}_{\pm}(x,\xi,t)$ is
\be
\int_0^1 dx \, x H_+^{(M)}(x,\xi,t)= 2 J^q(t)\,, \ \ \  \int_0^1 dx \,   H_-^{(M)}(x,\xi,t)= G^{(M) \, q}(t)\,,
\ee
where $J^q(t)$ is the ($t$-dependent) fraction of the nucleon angular momentum carried by a specific quark flavor, and $G^{(E,M) \, q}(t)$
are the usual electric and magnetic combinations of nucleon's form factors,
\be
G^{(E) \, q}(t)=F_1^q(t)+\tau F_2^{q}(t)\,, \quad G^{(M) \,q }(t)=F_1^{q}(t)+F_2^{q}(t)\,.
\ee

We introduce the electric and magnetic elementary LO amplitudes:
\begin{align}
  {\cal H}^{(E)}_\pm(\xi, t) & = \int_{-1}^1 d x\left(H^q (x, \xi, t)+ \tau E^q (x, \xi, t)\right)\left[\frac{1}{\xi-x-i 0} \mp \frac{1}{\xi+x-i 0}\right] \nn  \\
& = \int_0^1 d x H_{\pm}^{(E)}(x, \xi, t)\left[\frac{1}{\xi-x-i 0} \mp \frac{1}{\xi+x-i 0}\right];
\end{align}
and
\begin{align}
  {\cal H}^{(M)}_\pm(\xi, t) & = \int_{-1}^1 d x\left(H^q (x, \xi, t)+E^q (x, \xi, t)\right)\left[\frac{1}{\xi-x-i 0} \mp \frac{1}{\xi+x-i 0}\right] \nn  \\
 & = \int_0^1 d x H_{\pm}^{(M)}(x, \xi, t)\left[\frac{1}{\xi-x-i 0} \mp \frac{1}{\xi+x-i 0}\right]\,,
\end{align}
with
\be
\im \mathcal{H}_{\pm}^{(E,M)}(\xi, t)= \pi H_{\pm}^{(E,M)}(\xi, \xi, t)\,.
\ee

In the following, we focus on the case of the singlet combinations of electric and magnetic GPDs
$H_+^{(E,M)}$, which is of practical  importance for the description of DVCS.
The electric singlet elementary amplitude (Compton FF) satisfies the once subtracted fixed-$t$ dispersion relation analogous to (\ref{DispRel_start}):
\be
\operatorname{Re} \mathcal{H}_{+}^{(E)}(\xi, t)=\mathcal{P} \int_0^1 d x \frac{2 x H_{+}^{(E)}(x, x, t)}{\xi^2-x^2}+4 (1-\tau) D(t)\,,
\label{DispRel_E}
\ee
and for the magnetic singlet Compton FF the once subtracted fixed-$t$ dispersion relation
takes the form of
\be
\operatorname{Re} \mathcal{H}_{+}^{(M)}(\xi, t)=\mathcal{P} \int_0^1 d x \frac{2 x H_{+}^{(M)}(x, x, t)}{\xi^2-x^2}\,.
\label{DispRel_M}
\ee
The subtraction constant in the magnetic case is zero due to the exact cancellation between
the $D$-term contributions into GPDs $H$ and $E$.

In the present analysis we neglect the effects associated with the
target mass and set the relativistic velocity factor $\beta$
(\ref{Def_beta}) to $1$.
Working out the Froissart-Gribov projection for the electric combination fully repeats that for the case of spinless target. The cross-channel SO$(3)$ PWs are defined as in Eq.~(\ref{Def_FJ_FF}).
Therefore,  for even $J \ge 0$
\be
F_{J=0}^{(E)}(t)=2 \int_0^1 d x\left[\frac{\mathcal{Q}_0(1 / x)}{x^2}- \frac{1}{x}\right] H_{+}^{(E)}(x, x, t)+4 (1- \tau) D(t)\,,
\label{FJ0_electric}
\ee
and
\be
F_{J>0}^{(E)}(t)=2 (2J+1)\int_0^1 d x \frac{\mathcal{Q}_0(1 / x)}{x^2} H_{+}^{(E)}(x, x, t)\,.
\label{FJ_electric}
\ee

The magnetic case is a bit more intricate.
The corresponding cross channel SO(3)-PWs are defined for even $J \ge 2$ as
\be
&&
F_J^{(M)}(t)=   \frac{2J+1}{2J(J+1)} \int_{-1}^1 d ( \cos \theta_t) {\cal H}^{(M)}( \cos \theta_t,t) \frac{1}{ \cos \theta_t} \left(1- { (\cos \theta_t)^2} \right) C_{J-1}^{3/2}( \cos \theta_t)\,.
\label{Def_PWs_magnetic}
\nn
\ee
This definition is consistent with the expansion of the conformal moments of the magnetic combination of nucleon GPDs
in
$
\frac{1}{\xi} P_J^{\prime}\left(\frac{1}{\xi}\right)= \frac{1}{\xi} C_{J-1}^{\frac{3}{2}} \left( \frac{1}{\xi} \right)
$
within the dual parametrization framework, {\it cf.} Eq.~(\ref{PW_dual_Magnetic}).

\begin{figure}[H]
 \begin{center}
\includegraphics[width=0.4\textwidth]{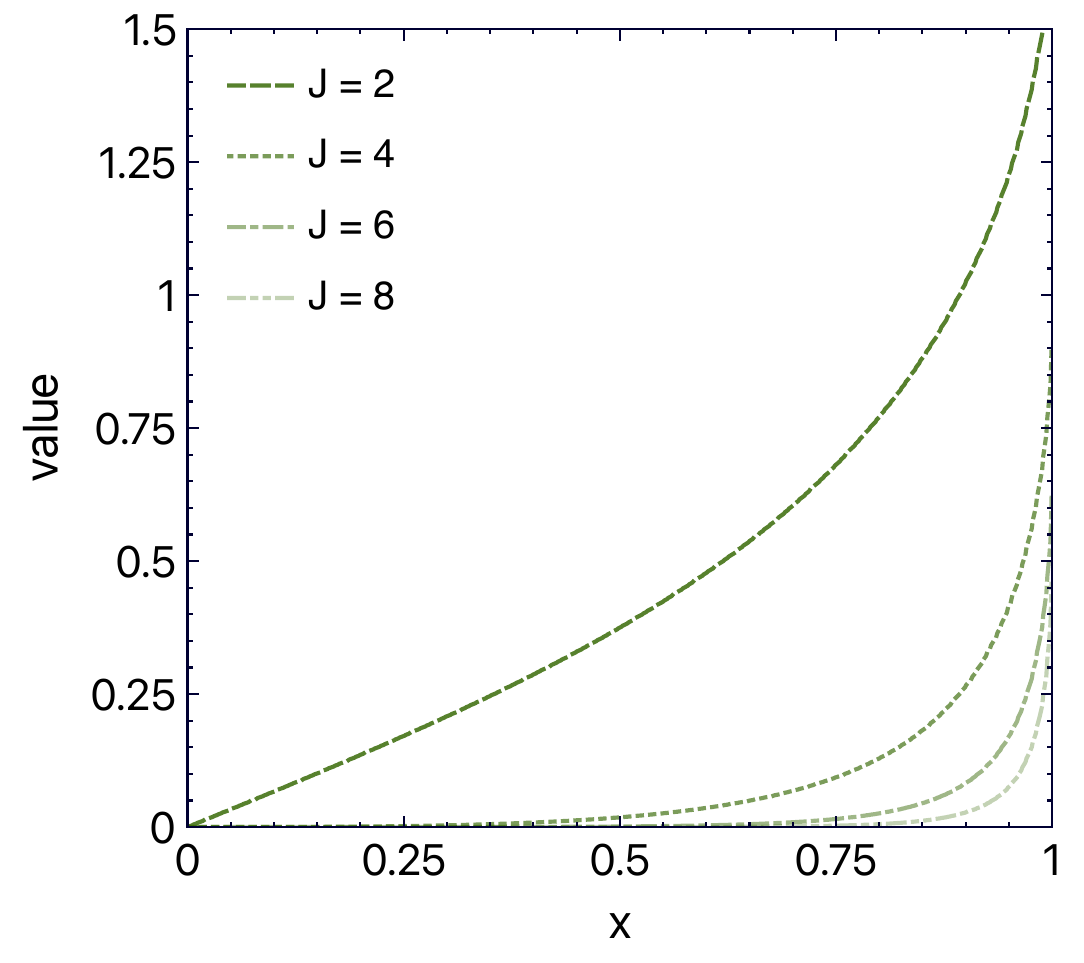}
\caption{First weight functions
$2 \frac{ 2J+1}{ J(J+1)} \frac{(-1)}{x}   \sqrt{\frac{1}{x^2}-1}  \,
{\cal Q}_J^1(1/x )$ entering the definition of the $F^{(M)}_J(t)$ projections (\ref{FG_projection_Magnetic}).}
 \label{Fig_Q_functions_M1}
  \end{center}
\end{figure}

To work out the FG projection we employ the dispersive representation (\ref{DispRel_M})
analytically continued to the $t$-channel
\be
{\cal H}^{(M)}( \cos \theta_t,t) = \int_0^1 dx \frac{2x (\cos \theta_t)^2}{1 -x^2 (\cos \theta_t)^2} H_+^{(M)}(x,x,t)
\ee
and plug it into the definition (\ref{Def_PWs_magnetic})
\begin{align}
F_J^{(M)}(t) & = \frac{2J+1}{2J(J+1)} \int_{-1}^1 d ( \cos \theta_t) {\cal H}^{(M)}( \cos \theta_t) \frac{1}{ \cos \theta_t} (-1) \left(1- { (\cos \theta_t)^2} \right)^{\frac{1}{2}}P_J^1( \cos \theta_t)
\nn \\
& = \int_0^1 dx  H_+^{(M)}(x,x,t)    \frac{2J+1}{2J(J+1)} \frac{(-1)}{x}  \int_{-1}^1 d ( \cos \theta_t)
\left(1- { (\cos \theta_t)^2} \right)^{\frac{1}{2}}P_J^1( \cos \theta_t)
\left( \frac{1}{\frac{1}{x} -  \cos \theta_t}+  \frac{1}{-\frac{1}{x}-  \cos \theta_t} \right)\,,
\label{Integral_to_compute_magnetic}
\end{align}
where we make use of the familiar relation between the Gegenbauer polynomials and the associated Legendre polynomials
\be
C^{\frac{3}{2}}_{J-1}(x)=(-1) (1-x^2)^{-\frac{1}{2}} P_J^1(x)\,.
\ee
The $\cos \theta_t$ integral in
(\ref{Integral_to_compute_magnetic})
can be performed using%
\footnote{See the integral {\bf 7.124} of Gradshteyn and Ryzhik \cite{Gradshteyn:1702455} for $k=0$.}:
\be
\int_{-1}^1  d y  \, (z-y)^{-1}\left(1-y^2\right)^{\frac{1}{2} m}  {P}_n^m(y)=2 \left(z^2-1\right)^{\frac{1}{2} m} { \cal Q}_n^m(z)\,,
\ee
where $ { \cal Q}_n^m(z)$ stands for the associated Legendre function of the second kind, which is defined
as
$$
 {\cal Q}_n^m(z)=\left(z^2-1\right)^{\frac{m}{2}} \frac{d^m  {\cal Q}_n(z)}{d z^m}
$$
in the complex plane with a cut along $(-1;\,1)$; and $m \le n$.
This finally yields for even $J \ge 2$
\be
F_J^{(M)}(t)
= 2 \int_0^1 dx  H_+^{(M)}(x,x,t)    \frac{ 2J+1}{ J(J+1)} \frac{(-1)}{x}   \sqrt{\frac{1}{x^2}-1}  \,
{\cal Q}_J^1(1/x )\,;
\label{FG_projection_Magnetic}
\ee
and the $J=0$ projection, naturally, makes no sense for the magnetic combination.
Figure~\ref{Fig_Q_functions_M1} shows several first weight functions
$2 \frac{ 2J+1}{ J(J+1)} \frac{(-1)}{x}   \sqrt{\frac{1}{x^2}-1}  \,{\cal Q}_J^1(1/x )$ entering definition of the $F^{(M)}_J(t)$ projections (\ref{FG_projection_Magnetic}).
Similarly to the FG of the electric GPD, the contribution of small-$x_B$ region is progressively
damped with the growth of $J$.

Since the dual parametrization framework is instrumental for the physical interpretation
of the FG projections of the Compton FFs, it is worth to present a brief summary
of the formalism for the spin-$\nicefrac{1}{2}$ target case.
The double PW expansion for $H_+^{(E,M)}$  reads
\be
H_{+}^{(E)}(x, \xi, t)=2 \sum_{\substack{n=1 \\ \text { odd even }}}^{\infty} \sum_{\substack{l=0 \\ \text { even }}}^{n+1} B_{n l}^{(E)}(t) \theta\left(1-\frac{x^2}{\xi^2}\right)\left(1-\frac{x^2}{\xi^2}\right) C_n^{\frac{3}{2}}\left(\frac{x}{\xi}\right) P_l\left(\frac{1}{\xi}\right)
\label{PW_dual_Electric}
\ee
and
\be
H_{+}^{(M)}(x, \xi, t)=2 \sum_{\substack{n=1 \\ \text { odd }}}^{\infty} \sum_{\substack{l=0 \\ \text { even }}}^{n+1} B_{n l}^{(M)}(t) \theta\left(1-\frac{x^2}{\xi^2}\right)\left(1-\frac{x^2}{\xi^2}\right) C_n^{\frac{3}{2}}\left(\frac{x}{\xi}\right) \frac{1}{\xi} P_l^{\prime}\left(\frac{1}{\xi}\right)\,,
\label{PW_dual_Magnetic}
\ee
where, in complete analogy with Eq.~(\ref{Def_Bnl}), the electric and and magnetic generalized form factors
are generated as the Mellin moments of corresponding forward-like functions
\be
B_{n, n+1-2 \nu}^{(E,M)}(t)=\int_0^1 d y y^n Q_{2 \nu}^{(E,M)}(y, t) \quad \text { or } \quad B_{n, l}^{(E,M)}(t)=\int_0^1 d y y^n Q_{n+1-l}^{(E,M)}(y, t)\,.
\ee
The forward-like functions $Q_0^{(E,M)}(y,t)$ are fixed in terms of $t$-dependent
quark densities $q_{+}(y, t)$ and $e_{+}(y, t)\equiv E_{+}(y, 0, t)$:
\be
&&
Q_0^{(E)}(y, t)=\left[\left(q_{+}(y, t)+\tau e_{+}^q(y, t)\right)-\frac{x}{2} \int_y^1 \frac{d z}{z^2}\left(q_{+}(z, t)+\tau e_{+}^q(z, t)\right)\right]\,, \nn \\ &&
Q_0^{(M)}(y, t)=\frac{1}{2} \int_y^1 \frac{d z}{z}\left(q_{+}(z, t)+e_{+}^q(z, t)\right)\left[1+\frac{y}{z}\right]\,.
\ee
The normalization is provided by the momentum and angular
momentum sum rules:
\be
\begin{aligned}
B_{1,2}^{(E)}(t) & \equiv \int_0^1 d x x Q_0^{(E)}(x, t)  =\frac{5}{6}\left[M_2^q(t)(1-\tau)+\tau 2 J^q(t)\right]\,, \\
B_{1,2}^{(M)}(t) & \equiv \int_0^1 d x x Q_0^{(M)}(x, t)  =\frac{5}{12} 2 J^q(t)\,.
\end{aligned}
\ee

The Mellin moments of the GPDs
(\ref{PW_dual_Electric}),
(\ref{PW_dual_Magnetic})
for odd $N \ge 1$ read
\be
&&
\int_0^1 d x x^N H_{+}^{(E)}(x, \xi, t)=\sum_{\substack{k=0 \\ \text { even }}}^{N+1} h_{N, k}^{(E)}(t) \xi^k\,,
\nn \\ &&
\int_0^1 d x x^N H_{+}^{(M)}(x, \xi, t)=\sum_{\substack{k=0 \\ \text { even }}}^{N-1} h_{N, k}^{(M)}(t) \xi^k\,,
\ee
where the coefficients for the electric combination%
\footnote{{\it Cf.} Eq.~(\ref{FF_decomp_Diehl}) for the definition of the coefficients $ {\cal A}_{N+1,k}(t)$, ${\cal B}_{N+1,k}(t)$ and
$ {\cal C}_{N+1 }(t)$ through the form factor decomposition of nucleon matrix elements of
twist-$2$ local quark operators (\ref{Def_operator_twist2}).
},
\be
h_{N,k}^{(E)}(t) \equiv 2^k \left( {\cal A}_{N+1,k}(t) + \tau  {\cal B}_{N+1,k}(t) \right)+ \delta_{N+1,k} 2^{N+1} {\cal C}_{N+1}(t)  \ \ \ \text{with even} \ \ k \le N+1\,,
\label{Electric_h}
\ee
are expressed through the generalized FFs
$B_{n,l}^{(E)}(t)$
by Eq.~(\ref{Mellin_moments_coefs_through_Bnl}).
The coefficients for the magnetic combination,
\be
h_{N,k}^{(M)}(t) \equiv 2^k \left( {\cal A}_{N+1,k}(t) +    {\cal B}_{N+1,k}(t) \right)
 \ \ \text{with even} \ \ \ k \le N-1\,,
 \label{Magnetic_h}
\ee
are expressed through
$B_{n,l}^{(M)}(t)$
as
\begin{equation}
\begin{aligned}
& h_{N, k}^{(M)}(t)=\sum_{\substack{n=1 \\
\text { odd }}}^N \sum_{\substack{l=2 \\
\text { even }}}^{n+1} B_{n l}^{(M)}(t)(-1)^{\frac{k+l-N-1}{2}} \frac{\Gamma\left(1-\frac{k-l-N}{2}\right)}{2^k \Gamma\left(\frac{1}{2}+\frac{k+l-N}{2}\right) \Gamma(2-k+N)}
\frac{(1-k+N)(n+1)(n+2) \Gamma(N+1)}{\Gamma\left(1+\frac{N-n}{2}\right) \Gamma\left(\frac{5}{2}+\frac{N+n}{2}\right)}\,.
\end{aligned}
\end{equation}

The Abel transform tomography method of Ref.~\cite{Polyakov:2007rv} admits a straightforward generalization
for the electric and magnetic Compton FFs. This requires introducing the electric and magnetic
GPD quintessence functions
\be
N^{(E,M)}(y, t) \equiv \sum_{\nu=0}^{\infty} y^{2 \nu} Q_{2 \nu}^{(E,M)}(y, t)\,.
\ee
The inversion formula for the electric case repeats the spinless case result (\ref{N_function_Abel}), while for even non-negative $J$ we obtain the following relation  between the Mellin moments of the electric GPD quintessence function and the
FG projection
\be
&&
F_{J=0}^{(E)}(t)
=4 \operatorname{Reg} \int_0^1 \frac{d y}{y}\left(N^{(E)}(y, t)-Q_0^{(E)}(y, t)\right)=
4\sum_{\nu=1}^\infty B_{2 \nu-1, J}^{(E)} (t)\,,
\nn \\ &&
F^{(E)}_{J>0}(t)
=
4\int_0^1 d y y^{J-1} N^{(E)}(y, t) =  4\sum_{\nu=0}^\infty B_{J+2 \nu-1, J}^{(E)} (t)\,,
\label{FJ_through_N_Electric}
\ee
where, analogously to Eq.~(\ref{F_J=0_through_N}), the $J=0$ integral may require a regularization of the Regge-like
singularity for $y=0$.

For the magnetic combination the  function $N^{(M)}(y,t)$
is expressed through the inverse Abel transformation:
\be
N^{(M)}(y,t)= \frac{1}{2 \pi } \frac{1-y^2}{\sqrt{1+y^2}} \int_{\frac{2y}{1+y^2}}^1  \frac{d x}{\sqrt{x}} \frac{1}{\sqrt{x- \frac{2y}{1+y^2}}}  \im { \cal H}^{(M)}_+(x, t)\,.
\label{N_magnetic_Abel}
\ee
To check that the analogue of (\ref{FJ_through_N_Electric}) also holds for the magnetic combination and
for even $J> 0$,
\be
F^{(M)}_{J>0}(t)
=
4\int_0^1 d y y^{J-1} N^{(M)}(y, t) =  4\sum_{\nu=0}^\infty B_{J+2 \nu-1, J}^{(M)} (t)\,,
\label{FJ_through_N_Magnetic}
\ee
we compute the $J-1$-th Mellin moment of (\ref{N_magnetic_Abel})
\be
\int_0^1 dy \,y^{J-1} N^{(M)}(y,t)= \frac{1}{ \pi} \int_0^1  d x \,  \im {\cal H}^{(M)}_+(x,t) \, \frac{1}{\sqrt{2 x}} \int_0^x \frac{dw}{w^{\frac{3}{2}}}
\left( \frac{1-\sqrt{1-w^2}}{w} \right)^{J+ \frac{1}{2}} \frac{1}{\sqrt{x-w}}\,,
\label{RJmagnetic_integral}
\ee
where we interchanged the order of integration and performed the integration variable substitution
$\frac{1}{w}= \frac{1}{2} \left(y+ \frac{1}{y} \right)$ corresponding to the Joukowski conformal map.
One may verify that indeed the $w$-integral in
(\ref{RJmagnetic_integral})
provides the desired result
\be
\frac{1}{\sqrt{2 x}} \int_0^x \frac{dw}{w^{\frac{3}{2}}}
\left( \frac{1-\sqrt{1-w^2}}{w} \right)^{J+ \frac{1}{2}} \frac{1}{\sqrt{x-w}}= - \frac{J+ \frac{1}{2}}{J(J+1)} \frac{1}{x}
%
\left[
\sqrt{ \frac{1}{x^2}-1} {\cal Q}_J^1 \left( 1/x \right)
\right]\,,
\ee
that corresponds to the weight function in the FG projection
(\ref{FG_projection_Magnetic}).

Analogously to the spinless target case, by truncating the $\nu$-summation in Eqs.~(\ref{FJ_through_N_Electric}), (\ref{FJ_through_N_Magnetic})
one may establish a set of sum rules relating the electric and magnetic
spin-$J$ FG projections to the combinations of the
coefficients at powers of $\xi$ of the Mellin moments of GPDs
(\ref{Electric_h}),
(\ref{Magnetic_h}), which can be expressed in terms
of the form factors of the nucleon matrix elements (\ref{FF_decomp_Diehl}) of local twist-$2$ operators (\ref{Def_operator_twist2}).

The resulting sum rules for the electric FG projections coincide with those
established in Sect.~\ref{Sec_Spinless_case} for the spinless case (see Eqs. (\ref{FJ=0_SR}) and (\ref{FJ=2_SR}) for $J=0,2$ sum rules), with the obvious replacement
$h_{N,k}(t) \to h_{N,k}^{(E)}(t)$.
Here, as an example, we would like to consider the $J=2$ magnetic FG projection
with truncation at $\nu_{\max}=1$.
It results in the following sum rule
\begin{align}
F_{J=2}^{(M)}(t) & = 4 \left( B_{1,2}^{(M)}(t)+B_{3,2}^{(M)}(t) + \ldots \right)  \nn \\
& =
-\frac{7}{12} h_{1,0}^{(M)}(t)+
\frac{9}{4}  h_{3,0}^{(M)}(t)
+\frac{21}{4}
 h_{3,2}^{(M)}(t)  +\genfrac\{\}{0pt}{0}{\text{contribution of conformal PWs}}{\text{with} \; \nu \ge 2}\,,
\end{align}
where the normalization of $h_{1,0}^{(M)}(t)$, $h_{3,0}^{(M)}(t)$
is fixed from the $\xi \to 0$ limit of the magnetic nucleon GPD:
\be
h_{1,0}^{(M)}(t) \equiv \int_0^1 dx x (q_+(x,t)+e_+(x,t))=2 J^q(t)\,, 
\ \ \ \
h_{3,0}^{(M)}(t) \equiv \int_0^1 dx x^3 (q_+(x,t)+e_+(x,t))\,.
\ee

\section{Froissart-Gribov of the Compton FFs: GPD models v.s. experiential data}
\label{Sec_Experiment_and_Models}

\begin{figure}[H]
\begin{center}
\includegraphics[width=0.30\textwidth]{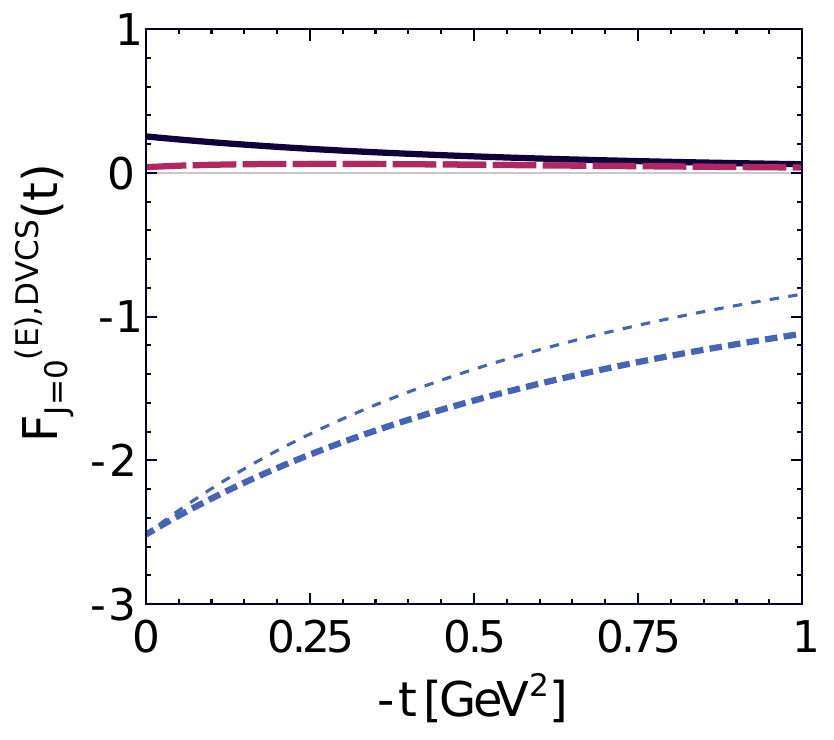}
\includegraphics[width=0.30\textwidth]{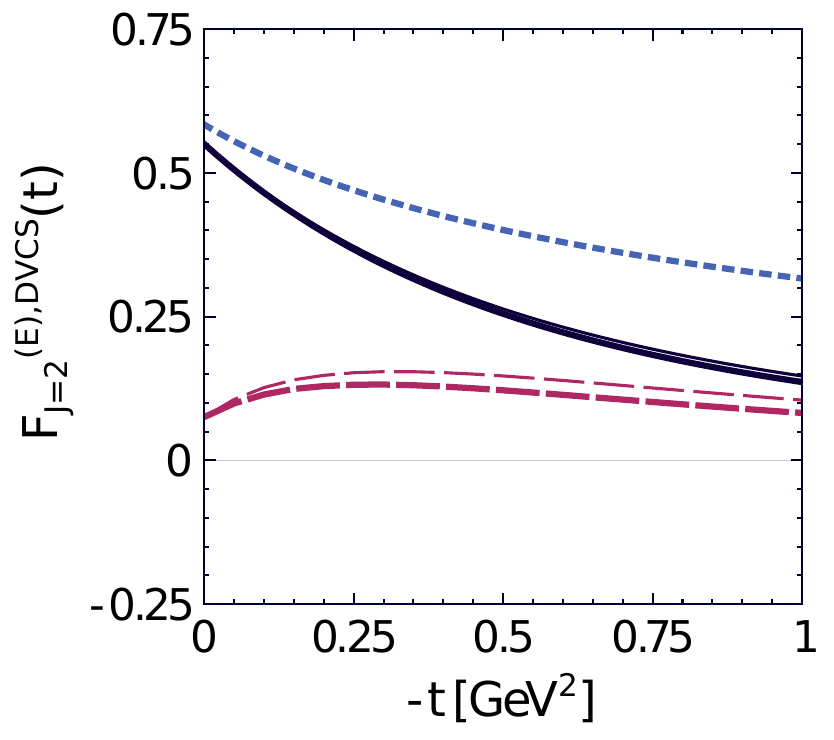}
\includegraphics[width=0.30\textwidth]{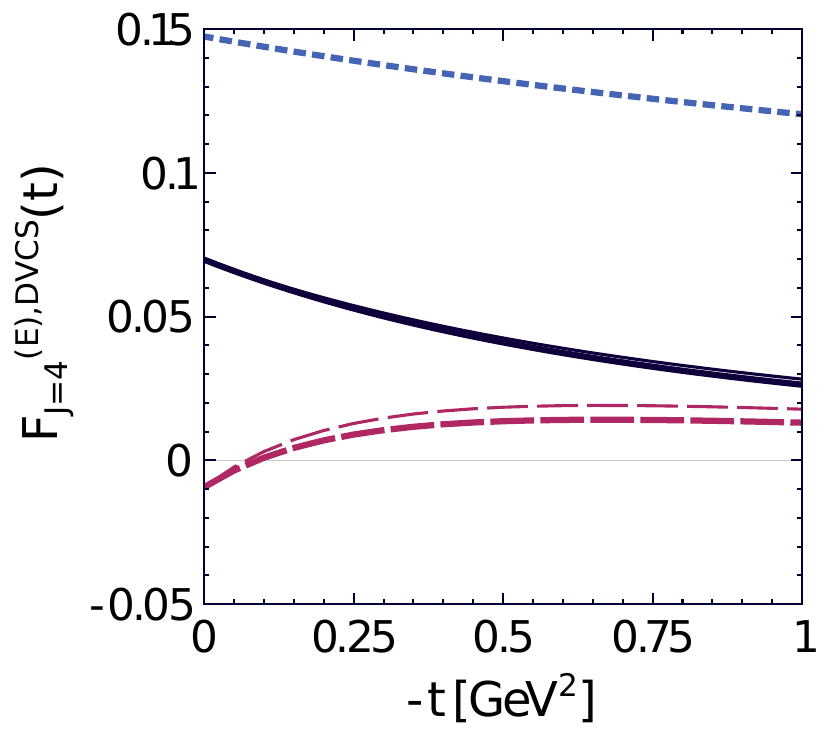}
\includegraphics[width=0.30\textwidth]{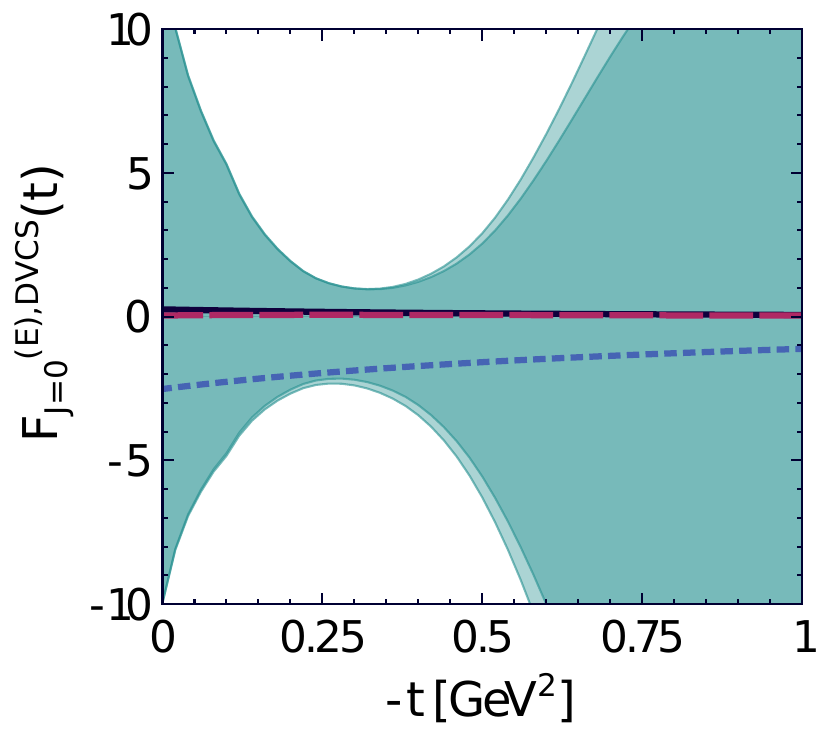}
\includegraphics[width=0.30\textwidth]{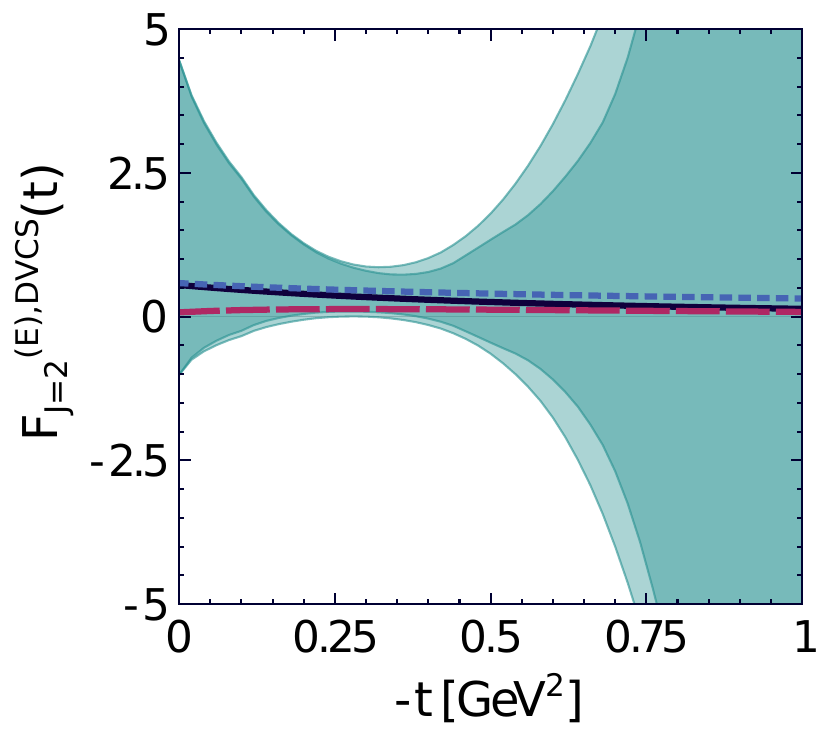}
\includegraphics[width=0.30\textwidth]{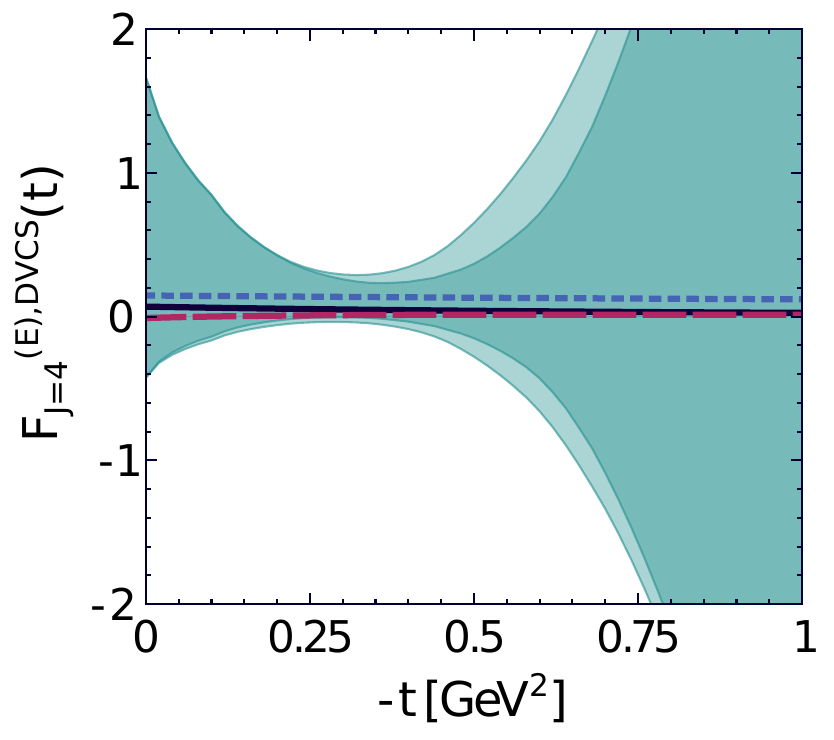}
\caption{
$F_{J}^{(E),\mathrm{DVCS}}$ for $J=0,2,4$ at $Q^2 = 2\,\mathrm{GeV}^2$ as a function of $t$.  First row: results obtained with the GK (solid black), MMS (dashed red) and KM (dotted blue) models. Thin lines denote estimates obtained with only GPD $H$. Second row: as before, with addition of results obtained with the CFFs coming from global analysis of the DVCS data \cite{Moutarde:2019tqa} (light turquoise bands, corresponding to $68\%$ confidence level). Dark inner bands are for results obtained with only the CFF $\mathcal{H}$.}
\label{fig:modelsDVCSE}
\end{center}
\end{figure}

\begin{figure}[H]
\begin{center}
\includegraphics[width=0.30\textwidth]{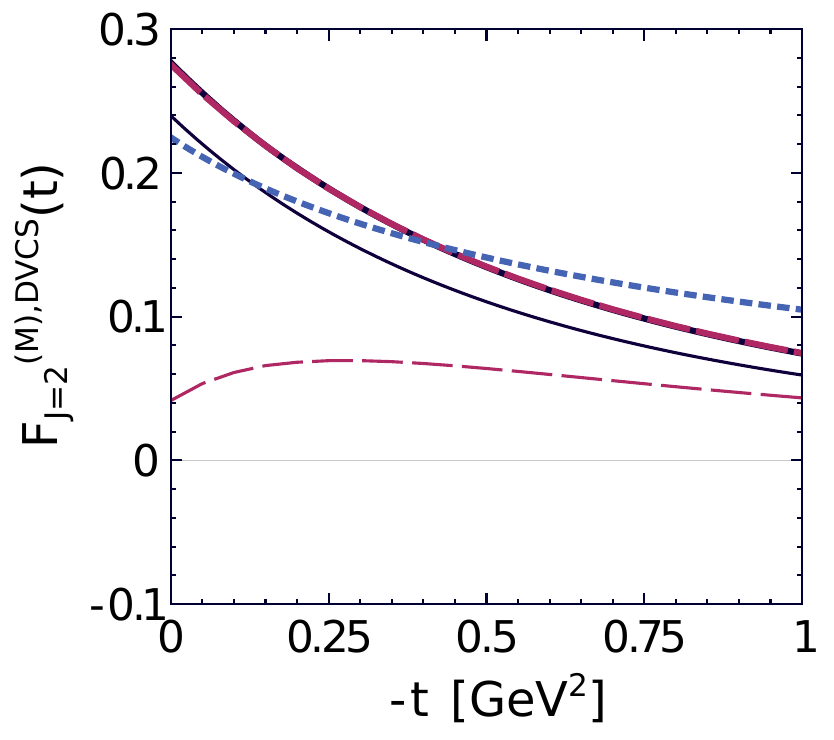}
\includegraphics[width=0.30\textwidth]{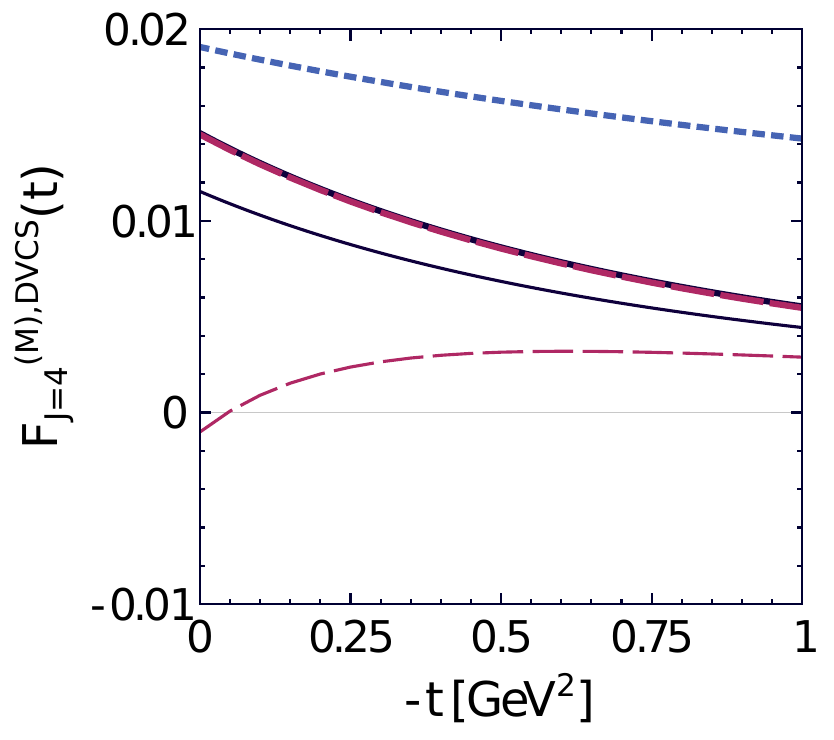}
\includegraphics[width=0.30\textwidth]{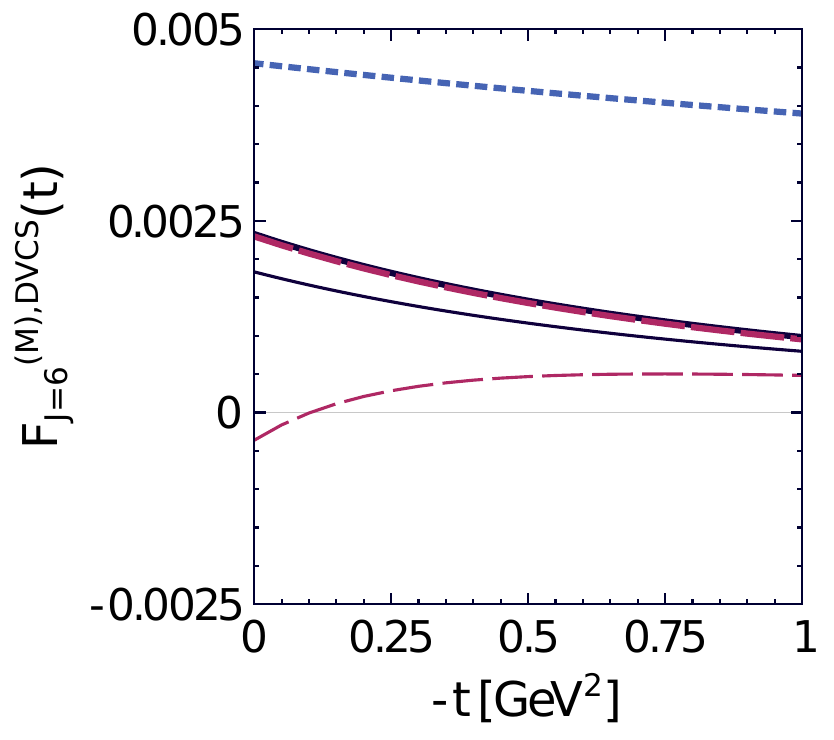}
\includegraphics[width=0.30\textwidth]{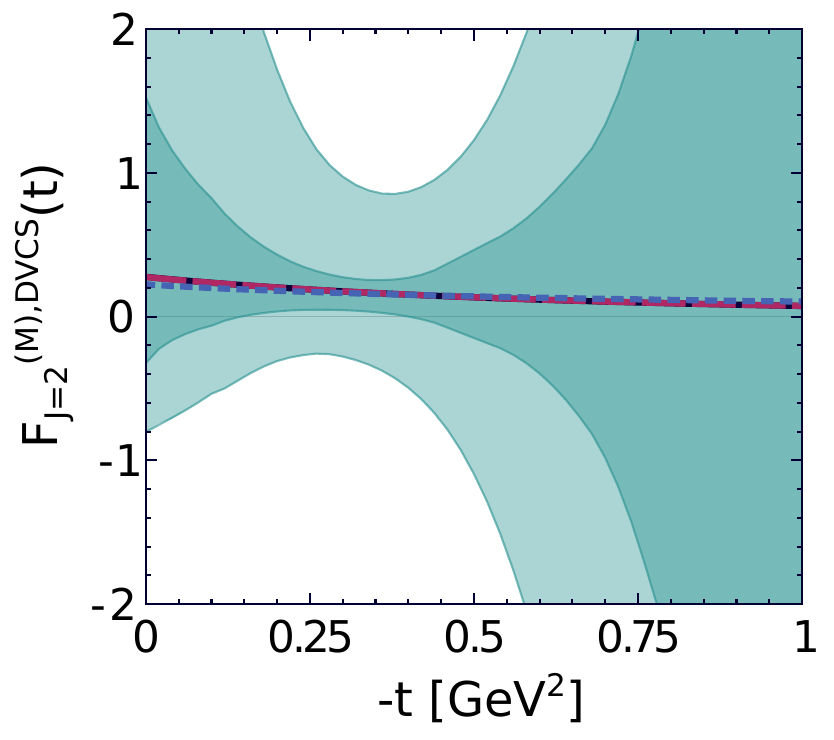}
\includegraphics[width=0.30\textwidth]{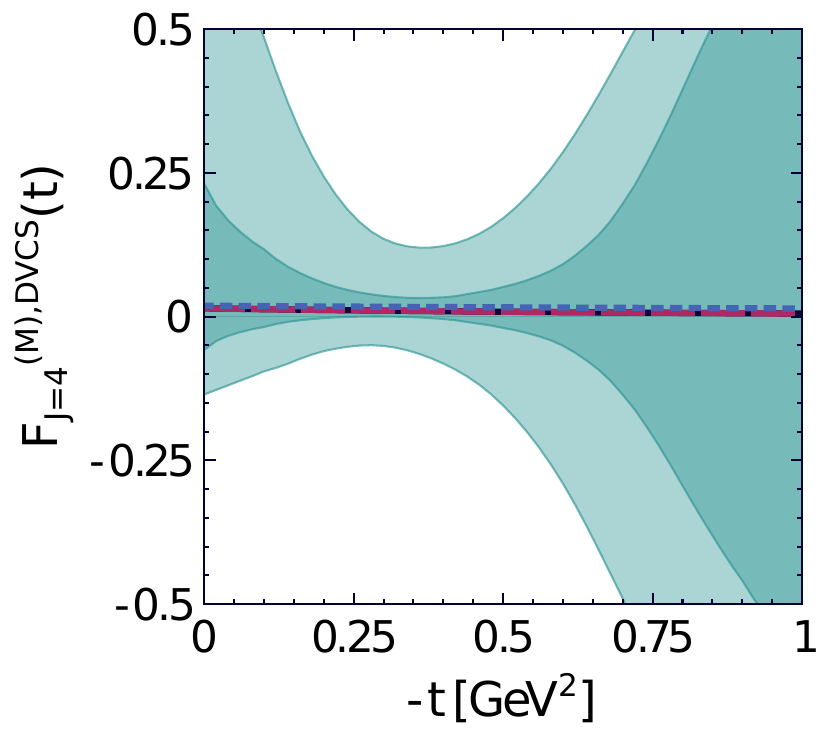}
\includegraphics[width=0.30\textwidth]{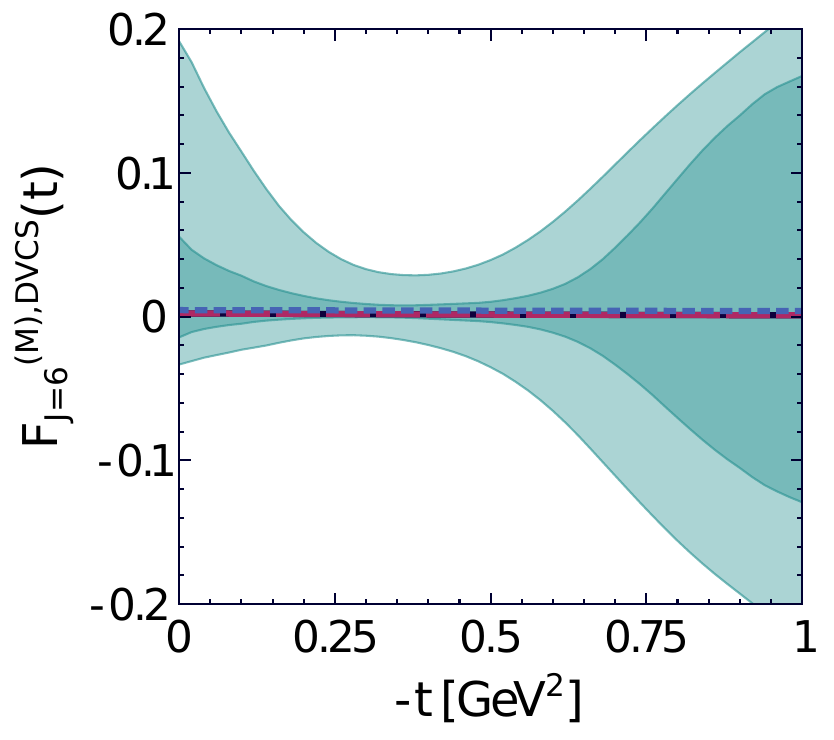}
\caption{$F_{J}^{(M),\mathrm{DVCS}}$ for $J=2,4,6$ at $Q^2 = 2\,\mathrm{GeV}^2$ as a function of $t$. For further description see the caption of Fig.~\ref{fig:modelsDVCSE}}
\label{fig:modelsDVCSM}
\end{center}
\end{figure}

In this Section we consider phenomenological application of the FG projections of CFFs.
The numerical calculations of a first few FG projections are presented in Fig.~\ref{fig:modelsDVCSE}
for
$F_{J}^{(E),\mathrm{DVCS}}$
and in
Fig.~\ref{fig:modelsDVCSM}
for
$F_{J}^{(M),\mathrm{DVCS}}$.
These results have been obtained with the GK \cite{Goloskokov:2005sd,Goloskokov:2008ib} and the MMS \cite{Mezrag:2013mya} models implemented in the PARTONS \cite{Berthou:2015oaw},  the KM15 model \cite{Kumericki:2015lhb} implemented in the GeParD
software package \cite{gepard}.
We also present estimates of the FG projections
accounting only the contribution of the best constrained nucleon GPD $H$.

We highlight substantial differences between GPD models used in our numerical analysis, helping us to illustrate sensitivity of the FG projections on the choice of the modelling strategy:
\bi
\item
The GK model is based on the so-called two-component  representation (double distribution (DD) part $+$ $D$-term) \cite{Polyakov:1999gs}; and has been primarily constrained by deeply vector meson production data collected at low-$x_B$.
\item The MMS model is a modified version of GK, where the valence parts of GPDs $H$ and $E$ have been replaced by new Ans{\"a}tze based on the one-component DD representation \cite{Radyushkin:2011dh}, with free parameters constrained by the DVCS measurements collected by the JLab Hall A and CLAS experiments.
\item The KM15 is a hybrid model with the sea contribution described with help of the double partial wave expansion, and valence one with dispersion integral representation. The model is constrained in a global analysis of the DVCS data.
\ei

We compare the results for the FG projections from these
phenomenological
GPD models to those computed from the global extraction of the  DVCS CFFs   obtained with  help of the machine learning technique relying on the artificial neural networks (ANNs). In Ref.~\cite{Moutarde:2019tqa} a set of simple ANNs
was used to represent the real and imaginary parts of the CFFs $\mathcal{H}$, $\mathcal{E}$, $\widetilde{\mathcal{H}}$ and $\widetilde{\mathcal{E}}$. The networks process the CFF kinematics, \emph{i.e.} values of $x_B$, $t$ and $Q^2$, and
the real and imaginary parts of the CFFs are extracted independently without imposing the dispersion relation. This allows to obtain a truly ``agnostic'' result enabling the examination of data quality and cross-checking the numerical procedure through verifying that the
resulting CFFs satisfy the appropriate dispersion relations.
The outcome of constraining the networks in a global analysis of the DVCS data is given as a set of replicas (for details of the replication process see Ref.~\cite{Moutarde:2019tqa}), allowing to propagate experimental uncertainties to other quantities, in the present case being the FG projections.

We start the discussion of results with $F_{J}^{(E),\mathrm{DVCS}}$ projections plotted in Fig.~\ref{fig:modelsDVCSE}. For $J=0$, the KM15 gives a substantially different result, as it is the only model from the presented set that includes a functional form of the $D$-term\footnote{Formally, $D$-term is included in the MMS, however, in Ref.~\cite{Mezrag:2013mya} the functional form of $t$-dependent normalisation factor of this quantity is not given, preventing us from using it in our numerical estimate.}. Addition of a realistic $D$-term (see for instance Ref.~\cite{Dutrieux:2021nlz}) in the GK and MMS models would lead to a similar picture as for the KM15, \emph{i.e.} the dominance of the
$4 (1-\tau) D(t)$
term in Eq.~\eqref{FJ0_electric}
over that involving
$H_{+}^{(E)}(x, x, t)$.
In Fig.~\ref{fig:modelsDVCSE} we also see that, despite the GK and MMS models have the same sea quark components, they result in
significantly different estimates for $F_{J}^{(E),\mathrm{DVCS}}$. This confirms that even the smallest-$J$ FG projections are mostly sensitive to the valence contribution. The estimates coming from the MMS are in general smaller in  magnitude than those from other models, which is a consequence of building this model with the one-component DD modelling scheme. Finally, results obtained with the CFFs directly extracted from experimental data show a window in $t$
reasonably well
constrained by current experiments, where the FG projections are measured with a satisfactory precision. As in this window $|\tau| \ll 1$, therefore GPD $E$, which is known with a poor precision, contributes only a little. We stress that future measurements, preferably at JLab due to the sensitivity of the FG projections to the valence contribution, should allow to distinguish between various GPD models.

The magnetic FG projections  $F_{J}^{(M),\mathrm{DVCS}}$
plotted in Fig.~\ref{fig:modelsDVCSM} are not sensitive to the $D$-term contribution, allowing for a straightforward comparison between all three GPD models for any value of $J$. It is remarkable, that the GK and MMS models give approximately the same estimates for $F_{J}^{(M),\mathrm{DVCS}}$, despite they present different contributions of GPD $H$ and already discussed results for $F_{J}^{(E),\mathrm{DVCS}}$. The explanation comes by looking at Eqs.~(103), (105), (106)
of Ref.~\cite{Mezrag:2013mya}, which suggest, that $H_+^{(M)}(x,\xi,t)$ combination is not sensitive to the differences between the one- and two-component DD modelling schemes. Since both the GK and MMS give essentially the same forward limits for GPDs $H$ and $E$, and employ the same profile functions, they give, in this case, very similar results. The measurement of the FG projections with CFFs directly extracted from experimental data displays much larger relative uncertainties than in the case of $F_{J}^{(E),\mathrm{DVCS}}$. This is caused by a limited knowledge of the contribution of GPD $E$, which here is not suppressed by the kinematic coefficient $\tau$.

\begin{figure}[H]
\begin{center}
\includegraphics[width=0.30\textwidth]{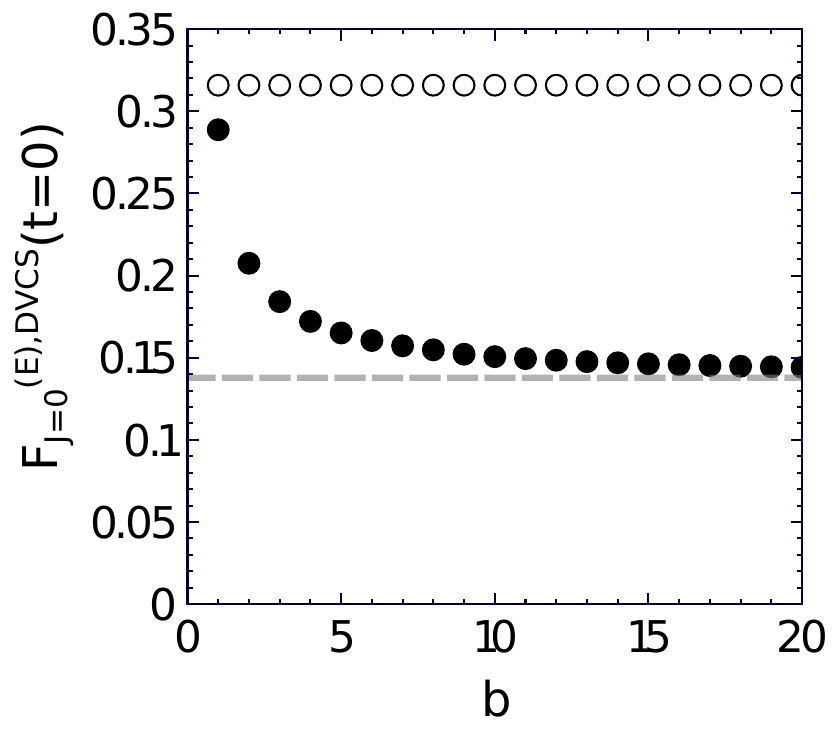}
\includegraphics[width=0.30\textwidth]{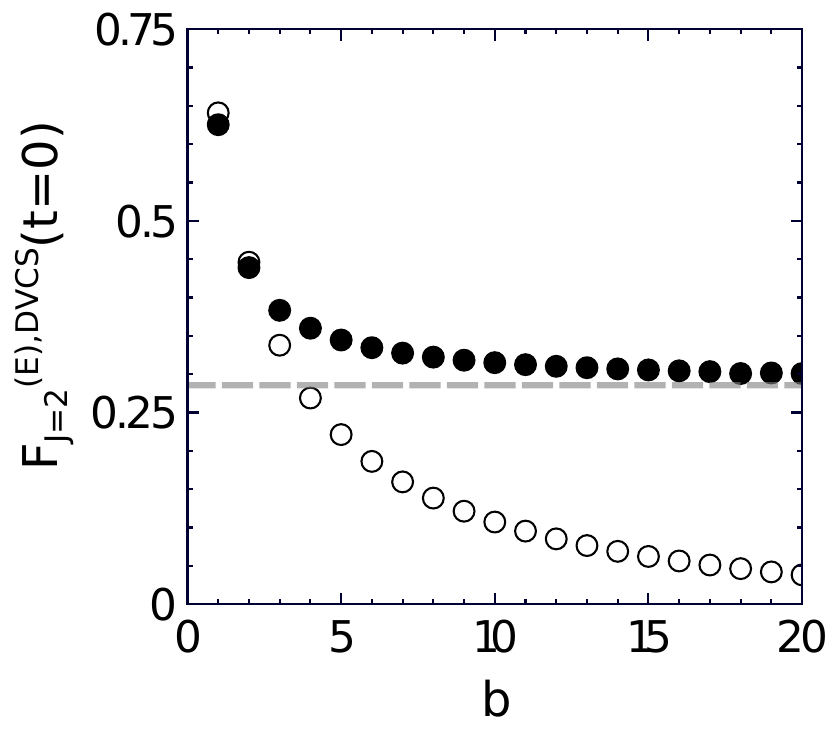}
\includegraphics[width=0.30\textwidth]{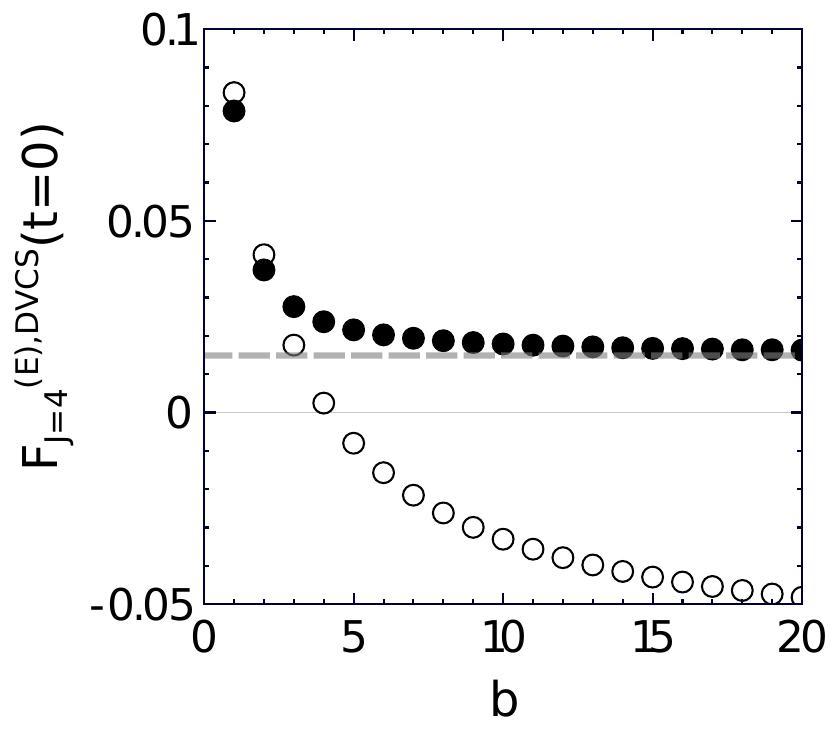}
\caption{$F_{J}^{(E),\mathrm{DVCS}}$ for $J=0,2,4$ at $t=0$ and $Q^2 = 2\,\mathrm{GeV}^2$ as a function of the double distribution profile parameter, $b$, simultaneously set for both valence and sea contributions. The GPD model used in this exercise is based on Eq.~\eqref{RDDA_b_dependent} and utilises the same PDFs as the GK model. Filled circles present the exact values for the FG projections, while open ones show the values obtained with the sum rules based on the ``next-to-minimalist'' dual model ($\nu_{\max} = 1$)
presented in Sect.~\ref{Sec_Spinless_case}. Note that the sum rule (\ref{FJ=0_SR}) for
$F_{J=0}^{(E),\mathrm{DVCS}}$ does not show any $b$-dependence since the coefficient
$h_{1,0}$
is fixed by the first Mellin of the parton density $q(x)$; and $h_{1,2}=0$ since the $D$-term is not included in (\ref{RDDA_b_dependent}).
The dashed lines denote the asymptotic result for the FG projections in the $b\to\infty$ limit.}
\label{fig:modelsProfile}
\end{center}
\end{figure}

Figure~\ref{fig:modelsProfile}  illustrates the sensitivity of the FG projections to the GPD modelling assumptions. The figure
shows
$F_{J}^{(E),\mathrm{DVCS}}(t=0)$ for $J=0,2,4$
as a function of the profile function parameter, $b$,  appearing in the commonly employed factorized Radyushkin's double distribution Ansatz (RDDA) \cite{Musatov:1999xp}:
\begin{equation}
H(x,\xi,t=0) = \int_{-1}^{1}d\beta\int_{-1+|\beta|}^{1-|\beta|}d\alpha\,\delta(\beta+\xi\alpha-x)
\,
q(\beta)
\frac{\Gamma(2b+2)}{2^{2b+1}\Gamma^2(b+1)}
\frac{\left((1-|\beta|)^2 - \alpha^2\right)^b}{(1-|\beta|)^{2b+1}}\,,
\label{RDDA_b_dependent}
\end{equation}
where
the normalisation of the profile function
ensures the correct reduction to the forward limit, $H(x,0,0) = q(x)$, and the parameter $b$ controls the buildup of the skewness effect under integration of a double distribution over the line $\alpha = (x-\beta)/\xi$. In particular,
$b\to\infty$
corresponds to vanishing skewness effect:
$H(x,\xi,0) = q(x)$.
In our exercise, we employ the same PDFs as in the  GK model and make use of the same value of $b$ both for valence and sea contributions. The obtained results demonstrate a strong sensitivity of the projections on the choice of this parameter. Such leverage may help in the future to constrain the parameter from experimental data.

In addition, in Fig.~\ref{fig:modelsProfile} we present a comparison between the exact evaluation of the FG projections $F_{J}^{(E),\mathrm{DVCS}}$ and values obtained relying on the sum rules based on the ``next-to-minimalist'' dual model ($\nu_{\max} = 1$)
presented in Sect.~\ref{Sec_Spinless_case}.
The exactly evaluated result for the FG projections for the double distribution model and the sum rules approximately agree only for small values of $b$. This behavior is well expected,
since, as demonstrated in Ref.~\cite{Polyakov:2009xir}, the reparametrization of a model
based on a $b$-dependent RDDA (\ref{RDDA_b_dependent}) into the dual parametrization framework requires to account
conformal PWs up to $\nu_{\max}=b-1$
in order to reproduce the skewness effect for small-$\xi$, that turns to be determining for the FG projections with small-$J$.
Therefore, a GPD model based on the RDDA with large-$b$ presents an example of a model for which the sum rules for the FG projections require to account a large number of subleading conformal partial waves. However, for small values of $b$, usually employed in phenomenology (\emph{e.g.} the GK model sets $b=1$ for valance and $b=2$ for sea quarks), the sum rules work reasonably well.
We conclude that a possible discrepancy between measured  $F_{J}^{(M),\mathrm{DVCS}}$ quantity and the sum rules, the latter being obtained for instance with the help of lattice QCD, may quantify the missing contribution of higher $\nu \ge 2$ subleading conformal partial waves, \emph{cf.} Eqs.~\eqref{F0_dual_parametrization} and~\eqref{FJ_dual_parametrization}. Such observation would be also interesting for studying target mass effects, see the discussion in Sect.~\ref{Sec_Spinless_case}, as contributions associated to higher $\nu$ are more prone to the mixing of PWs.

\section{Conclusions and Outlook}
\label{Sec_Conclusions}

The Froissart-Gribov projections of generalized Compton Form Factors arise
within the dispersive analysis of the DVCS
amplitudes expanded in the partial waves of the $t$-channel.
The FG projections are computed from the known
absorptive  parts of the CFFs through convolutions with the second kind Legendre functions. Since the fixed-$t$ dispersion
relation for the charge-even CFF  only  requires one subtraction, the  corresponding $J=0$ FG projection
includes the
explicit contribution from the $D$-term form factor.

 In this paper, we presented an overview of application of the FG projections in the context of DVCS and constructed a
proper generalization of the formalism for the case of spin-$\nicefrac{1}{2}$ target hadrons specifying the
projections for the electric, ${\cal H}^{(E)}={\cal H} + \tau {\cal E}$, and magnetic, ${\cal H}^{(M)}={\cal H} + {\cal E}$,
combinations of nucleon CFFs suitable for expansion in the cross channel SO$(3)$ PWs. We worked out the explicit expression for the FG projection of the magnetic combination of nucleon CFFs involving the associated Legendre functions of the second kind ${\cal Q}_J^1(1/x)$.

It is worth emphasizing that the FG projections represent observable quantities. Their calculation only requires a knowledge of the absorptive  parts of CFFs and the $D$-term form factor in the physical domain of the DVCS reaction. Moreover, it does not necessarily imply the analytic continuation in $t$ to the physical domain of the cross channel.

We presented the calculation of the $J=0,\,2,\,4$ FG projections of both electric and magnetic combinations
${\cal H}^{(E,M)}$
of the CFFs using the
Goloskokov-Kroll,
Kumeri\v{c}ki-M\"{u}ller,  and
Mezrag-Moutarde-Sabati\'e
phenomenological GPD models presently employed
for the analysis of the DVCS data.
We also compare the model results for the FG projections with those obtained from the
global model-independent extraction \cite{Moutarde:2019tqa} of the DVCS CFFs with use of the Artificial Neural Network
framework.
The FG projections turn to be rather discriminative
with respect to the phenomenological GPD models.
In particular,
we demonstrated the sensitivity of the FG projections to the parameter $b$ of the profile function of the commonly used Radyushkin's DD Ansatz.
Moreover, the projections with high $J$ mostly receive contributions from the large-$x_B$ domain and can be used to study the buildup of skewness effect at $\xi \to 1$.

A particularly clear interpretation of the FG projections is obtained within the dual parametrization of GPDs
based on the double (conformal and cross-channel SO$(3)$) PW expansion of GPDs.  The framework of the dual parametrization
allows an independent derivation of the FG projections in terms of the Mellin moments of the GPD quintessence functions $N^{(E,M)}(y,t)$
recovered from the imaginary parts of CFFs by means of the inverse Abel transformation. This brings a connection between
the FG projections and the generalized FFs  $B_{n,l}(t)$ of the dual parametrization, which, in turn, are in one-to-one correspondence with the FFs of hadronic matrix elements of local twist-$2$ operators occurring in the calculation of the Mellin moments of GPDs.

Truncating the double PW expansion at a certain value $\nu_{\max}$ of the parameter $\nu$ (\ref{Def_nu_parameter}),
which
specifies the difference between the conformal spin and the cross channel angular momentum, results in a set of
sum rules for the FG projections. These sum rules can be tested with phenomenological models of GPDs and also bring a new connection between the experimental data and the lattice QCD calculation of the GPD Mellin moments.

The approach based on the dual parametrization of GPDs provides an interpretation of the FG projections as quantities characterizing hadron's response on the cross channel excitations  with angular momentum  $J$. Thus the FG projection may be seen as a tool to expand the non-local string-like QCD probe created by the hard subprocess of hard exclusive into a tower of local probes of spin-$J$.
Moreover, relying on the FG projections it might be possible to introduce a new type of observables,
``spin-$J$ radii'' of hadrons, defined from the $t=0$ slopes of the spin-$J$ FG projections.
These quantities may complement the broadly discussed charge and mass (gravitational) radii of hadrons
(see {\it e.g.} Refs.~\cite{Polyakov:2018zvc,Kharzeev:2021qkd})
and provide information on hadron's ``effective charge distribution'' accessible with a cross channel probe of particular
spin-$J$.

The method also possesses a potential for generalization for the non-diagonal DVCS and DVMP reactions sensitive to transition GPDs,
which recently got a revival of both theoretical \cite{Kroll:2022roq,Kim:2023yhp,Semenov-Tian-Shansky:2023bsy}
and experimental \cite{CLAS:2023akb} attention.
Study of exciting the nucleon resonances with QCD non-local probes may provide new tools for baryon spectroscopy through
focusing on the resonances excited by the cross channel probe with particular angular momentum $J$
and allow to study resonances that are only very weakly coupled to conventional elementary probes.

 Finally, we stressed that the the interpretation of the FG projections in terms of hadron's response on cross channel spin-$J$ excitation becomes challenging once taking into account the effect of the target mass corrections. This results in admixture of higher spin contributions which is soaring up with increasing of the number of accounted PWs  $\nu_{\max}$.
Taming the mixing may be possible keeping $\nu_{\max}$ not too large. The consistency of this hypothesis can be tested with help of set of sum rules for the FG projections.
On the other hand, the better understanding of the mixing requires further studies addressing the analytic structure of the FG projections in $t$ and careful workout of the analytic continuation of the relevant FFs between the physical domain of the DVCS and that of the cross channel counterpart reaction $\gamma^* \gamma \to h \bar{h}$, {\it e.g.} generalizing the dispersive technique of Ref.~\cite{Pasquini:2014vua}.

\section*{Acknowledgements}

The work of K.S. was supported by Basic Science Research Program through the National Research Foundation of Korea (NRF) funded by the Ministry of Education  RS-2023-00238703; and under Grant
No. NRF-2018R1A6A1A06024970 (Basic Science Research Program); and by the Foundation for the Advancement of Theoretical Physics and Mathematics ``BASIS''. Furthermore, the authors would like to also thank  Ho-Meoyng Choi, Krzysztof Cichy, Chueng-Ryong Ji, Bernard Pire, Hyeon-Dong Son, Sangyeong Son, Lech Szymanowski, Marc Vanderhaeghen and Christian Weiss for
enlightening discussions.

\renewcommand*{\thesection}{\Alph{section}}

\appendix

\setcounter{section}{0}
\setcounter{equation}{0}
\renewcommand{\thesection}{\Alph{section}}
\renewcommand{\theequation}{\thesection\arabic{equation}}

\section{Case of charge-odd GPD $H_{-}$}
\label{App_C_odd}

In this Appendix we briefly  discuss the case of charge-odd GPD $H_{-}$, which
can be experimentally accessed  \emph{e.g.} through the diphoton photoproduction \cite{Grocholski:2022rqj}.
We deal with the SO$(3)$ partial waves in $\cos \theta_t$ with odd spin, rather than even spin,
and define the cross channel PWs of the elementary amplitude $\mathcal{H}_{-}$
(\ref{Def_Elem_Ampl})
as
\be
F_J(t)= \frac{(-1)^J (2J+1)}{2} \int_{-1}^1 d (\cos \theta_t) P_J(\cos \theta_t)  \mathcal{H}_-(\cos \theta_t, t)\,.
\label{a_C_odd}
\ee
By plugging  the unsubtracted dispersion relation
\be
\mathcal{H}_{-}(\cos \theta_t, t)= \int_0^1 dx \frac{2 \cos \theta_t}{-1+x^2 \cos^2 \theta_t } H_{-}(x,x,t)
\label{DR_C_odd}
\ee
into (\ref{a_C_odd}) we obtain the Froissart-Gribov projection for all odd $J \ge 1$:
\be
F_J(t)
=2 (2J+1) \int_0^1 dx \frac{{\cal Q}_J(1/x)}{x^2} H_{-}(x,x,t)=4\sum_{\nu=0}^\infty B_{J+2 \nu-1, J} (t)\,.
\label{F_J_for_odd_J}
\ee
Here the charge-odd generalized FFs $B_{nl}(t)$ of the dual parametrization framework
defined in the double PW expansion
\be
H_{-}(x, \xi, t)=2 \sum_{\substack{n=0 \\ \text { even }}}^{\infty} \sum_{\substack{l=1 \\ \text { odd }}}^{n+1} B_{n l}(t)\, \theta\left(1-\frac{x^2}{\xi^2}\right)\left(1-\frac{x^2}{\xi^2}\right) C_n^{\frac{3}{2}}\left(\frac{x}{\xi}\right) P_l\left(\frac{1}{\xi}\right)
\label{Dual_Param_Cminus}
\ee
are
generated by the Mellin moments of the charge-odd
forward-like functions  $Q_{2 \nu}(y,t)$.
The charge-odd function
 $Q_0(y,t)$
is fixed in terms of $t$-dependent charge-odd combination of PDFs
(\ref{Def_forward_limit}):
\be
Q_0(y,t)=q_{-}(y, t)-\frac{y}{2} \int_y^1 \frac{d z}{z^2} q_{-}(z, t)\,,
\label{Def_Q0odd}
\ee
and normalized to the electromagnetic form factor
\be
B_{1,0}(t)= \int_0^1 dy \, Q_0(y,t)= \frac{3}{4} F^q(t)\,.
\ee

For even $N \ge 0$ the Mellin moments of the GPD
(\ref{Dual_Param_Cminus}) reads:
\be
&& \int_0^1 d x x^N H_{-}(x, \xi, t)=
\sum_{\substack{k=0 \\ \text { even }}}^{N} h_{N, k}(t) \xi^k \,.
\ee

By truncating the series in (\ref{F_J_for_odd_J}) at $\nu_{\max}=1$, we establish the following sum rule for $J=1$:
\be
&&
F_{J=1}(t)= 4 \left(  B_{0,1}(t)+ B_{2,1}(t)+ \dots \right)=
\frac{5}{4} h_{0,0}(t)+\frac{21}{4} h_{2,0}(t)+\frac{35}{4} h_{2,2}(t)+
\genfrac\{\}{0pt}{0}{\text{contribution of conformal PWs}}{\text{with} \; \nu \ge 2},
\nn \\ &&
\ee
where
\be
 h_{0,0}(t)= \int_0^1 dy q_-(y,t)= F^q(t); \ \ \  h_{2,0}(t)= \int_0^1 dy y^2 q_-(y,t)\,.
\ee

\section{A link to the double PW expansion in the Mellin-Barnes integral framework}
\label{App_conformal_moments}
\mbox

In this Appendix we briefly summarize the relation established in Ref.~\cite{Muller:2014wxa} between the double partial wave
expansion coefficients
$B_{n,l}(t)$ (\ref{Def_Bnl})
of the dual parametrization approach and the conformal moments of GPDs
defined according to the conventions
employed within the framework of Refs.~\cite{Mueller:2005ed,Kumericki:2015lhb}
based on the Mellin-Barnes integral technique.

The conformal moments of the GPD $H$
\be
H_n(\xi, t)=\int_{-1}^1 d x c_n(x, \xi) H^q(x, \xi, t)
\label{Def_Conf_moments_Dieter}
\ee
are defined with respect to the   conformal basis
\be
c_n(x, \xi)=\xi^n \frac{\Gamma\left(\frac{3}{2}\right) \Gamma(1+n)}{2^n \Gamma\left(\frac{3}{2}+n\right)} C_n^{\frac{3}{2}}\left(\frac{x}{\xi}\right)\,.
\label{Conf_basis_Dieter}
\ee
The basis
(\ref{Conf_basis_Dieter})
is normalized in way that in the forward limit $\xi \to 0$ it gives rise to the usual Mellin moments:
$\lim _{\xi \rightarrow 0} c_n(x, \xi)=x^n$.

The conformal moments (\ref{Def_Conf_moments_Dieter}) of the charge-even GPD are further expended over the basis
of the cross channel SO$(3)$ PWs
\be
H_n(\xi, t)=\sum_{\nu=0}^{(n+1) / 2} \eta^{2 \nu} H_{n, n+1-2 \nu}(t) \hat{d}^{n+1-2 \nu}_{00}(\xi)\,, \quad \text { for odd } n\,,
\ee
where $\hat{d}_{00}^J(\xi)$ stand for the reduced Wigner $d$-functions:
\be
\hat{d}_{00}^J(\xi)=\frac{\Gamma\left(\frac{1}{2}\right) \Gamma(1+J)}{2^J \Gamma\left(\frac{1}{2}+J\right)} \xi^J P_J\left(\frac{1}{\xi}\right)\,.
\ee
Therefore, accounting for the normalization issues, the two sets of double PW expansion coefficients are related as:
\be
H_{n, n+1-2 \nu}(t)=\frac{\Gamma(3+n) \Gamma\left(\frac{3}{2}+n-2 \nu\right)}{2^{2 \nu} \Gamma\left(\frac{5}{2}+n\right) \Gamma(2+n-2 \nu)} B_{n, n+1-2 \nu}(t)\,.
\label{BtoH_connection}
\ee
The set of the forward like functions of the dual parametrization $ Q_{2 \nu}(y, t)$ can be
reconstructed from the set of  coefficients
$H_{n, n+1-2 \nu}(t) $
by the usual  Mellin inversion formula, see {\it e.g.} \cite{Polyanin}:
\be
y^{2 \nu} Q_{2 \nu}(y, t)=\frac{1}{2 \pi i} \int_{c-i \infty}^{c+i \infty} d j y^{-j-1} \frac{2^{2 \nu} \Gamma \left( \frac{5}{2}+j+2 \nu \right) \Gamma(2+j)}{\Gamma(3+j+2 \nu) \Gamma \left( \frac{3}{2}+j\right)} H_{j+2 \nu, j+1}(t)\,.
\ee

\bibliography{Froissart_Gribov_Bib}

\end{document}